\documentclass[a4paper,11pt]{article}
\pdfoutput=1 % if your are submitting a pdflatex (i.e. if you have
\usepackage{jcappuba} % for details on the use of the package, please
\usepackage[T1]{fontenc} % if needed

\usepackage{amsmath}
\usepackage{amssymb}
\usepackage{amsfonts}
\usepackage{bm}
\usepackage{verbatim}
\usepackage{tablefootnote}
\usepackage{enumerate}
\usepackage{afterpage}
\usepackage{xcolor}
\usepackage{natbib, hyperref}

\usepackage{graphicx}
\usepackage{tabularx}
\usepackage{booktabs}
\usepackage{pifont}

\newcommand{\orcid}[1]{\href{https://orcid.org/#1}{\,\includegraphics[width=8px]{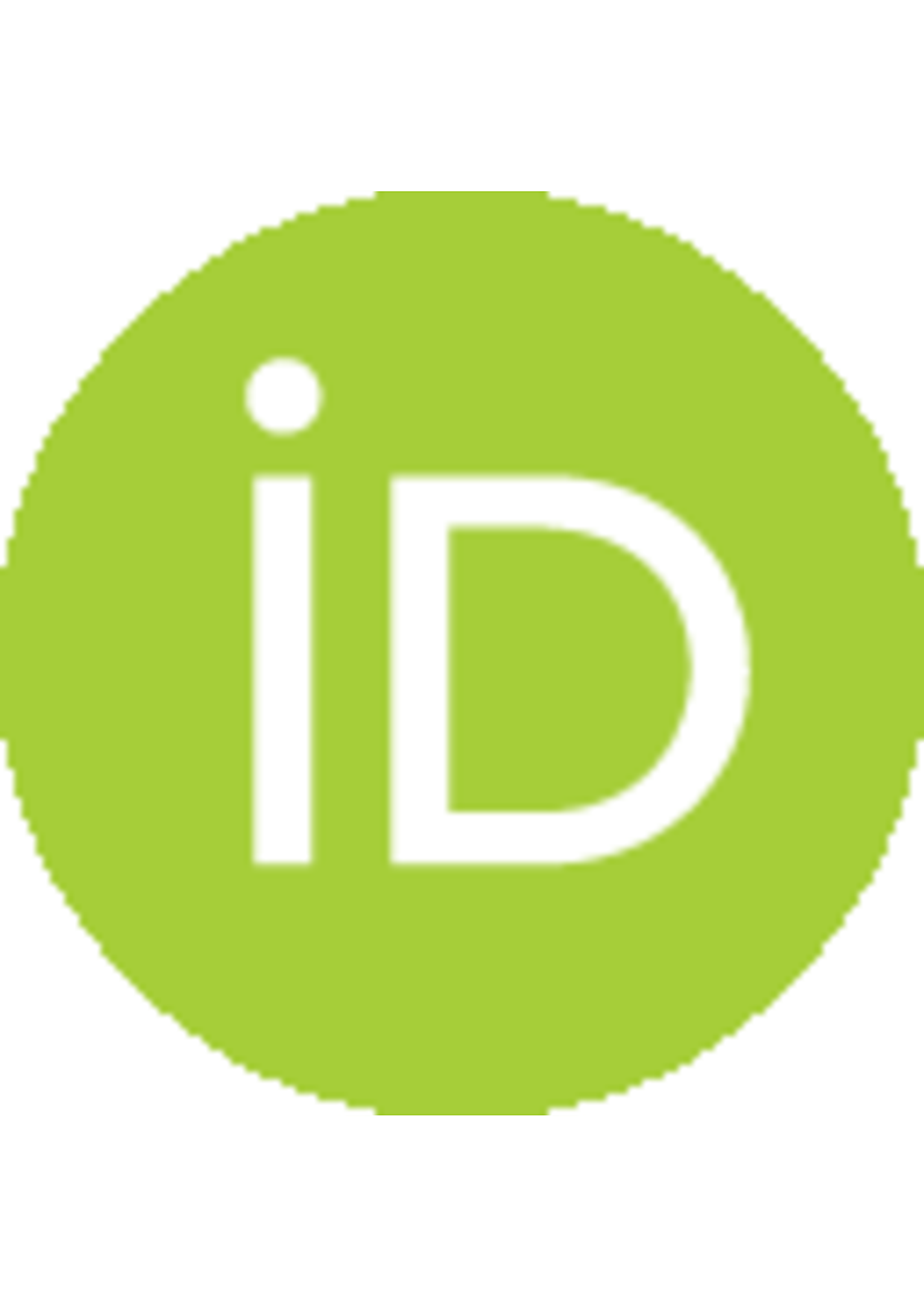}}}

\title{\boldmath Improved null tests of $\Lambda$CDM and FLRW in light of DESI DR2}

\author{Bikash R. Dinda$^{a}$,
Roy Maartens$^{a,b,c}$,
Shun Saito$^{d,e}$,\\
Chris Clarkson$^{f,a}$}

\affiliation[a]{Department of Physics \& Astronomy, University of the Western Cape, Cape Town 7535, South Africa}

\affiliation[b]{Institute of Cosmology \& Gravitation, University of Portsmouth, Portsmouth PO1 3FX, United Kingdom}

\affiliation[c]{National Institute for Theoretical \& Computational Science, Cape Town 7535, South Africa}
\affiliation[d]{Institute for Multi-messenger Astrophysics \& Cosmology, Department of Physics, Missouri University of Science \& Technology,  
Rolla, MO 65409, United States of America}
\affiliation[e]{Kavli Institute for the Physics \& Mathematics of the Universe, University of Tokyo, Kashiwa, Chiba 227-8583, Japan}
\affiliation[f]{Department of Physics \& Astronomy, Queen Mary University of London, London E1 4NS, United Kingdom}

\emailAdd{bikashrdinda@gmail.com}
\emailAdd{roy.maartens@gmail.com}
\emailAdd{saitos@mst.edu}
\emailAdd{chris.clarkson@qmul.ac.uk}

\abstract{
The DESI DR2 BAO data, in combination with CMB and different SNIa datasets, exclude the flat $\Lambda$CDM model at more than 2.5$\sigma$ when analyzed through the $w_0w_a$CDM parametrization for evolving dark energy. This simple parametrization may not accurately capture the behavior of the entire redshift range at late times, which may introduce bias in the results. We use null tests that probe for deviations from flat $\Lambda$CDM at late times, independent of any specific dark energy parametrization. We provide several diagnostics for null tests and discuss their advantages and disadvantages. In particular, we derive diagnostics that improve on previous ones, such as the popular $O_{\rm m}$ diagnostic. The diagnostics are derived from both background and perturbed quantities. Using the combination of DESI DR2 BAO and supernova data, with or without CMB data, we find that deviations from flat $\Lambda$CDM are at $\sim1\sigma$ confidence level in most of the redshift range (more than 1$\sigma$ for a few small redshift intervals in a few cases). When considering SDSS BAO data instead of DESI BAO data, in combination with PantheonPlus, with or without CMB data, we find even smaller deviations. Since spatial curvature can potentially modify the results, we also test for curvature in the general $\Lambda$CDM model and the general FLRW model. While there is slight evidence for nonzero cosmic curvature at lower redshifts in a general $\Lambda$CDM model, there is no statistically significant evidence in a general FLRW model.
}

\begin{document}
\maketitle
\flushbottom

\section{Introduction}

After the cosmological results of Dark Energy Spectroscopic Instrument (DESI) Data Release 1 (DR1), dynamical dark energy has gained a lot of attention, especially since there is evidence against the flat $\Lambda$CDM ($\Lambda$ Cold Dark Matter) model \citep{DESI:2024mwx}, from combined DESI DR1, cosmic microwave background (CMB) \citep{Planck:2018vyg} and type Ia supernovae (SNIa) \citep{Scolnic:2021amr} observations. The evidence against $\Lambda$CDM is slightly stronger with DESI DR2 data \cite{DESI:2025zgx}. This evidence is mostly based on the $w_0w_a$CDM parametrization and similar 2-parameter parametrizations of the dark energy equation of state.

Although these parametrizations are useful to test $\Lambda$CDM via data analysis, they may not capture the correct behavior of dark energy over a large redshift range. Consequently, they may introduce bias in the test and more parameters may be needed to reduce the bias (e.g. \citep{Nesseris:2025lke,Shlivko:2025fgv}). Evidence against $\Lambda$CDM  should be studied through physically motivated dark energy models, e.g. quintessence \citep{Caldwell:1997ii,Caldwell:2005tm,Tsujikawa:2013fta} (with better or equivalent evidence to fit the data) or through model-agnostic approaches.

Among the different model-agnostic approaches, a simple but efficient one is the construction of consistency, or null, tests. 
The basic idea is to construct functions of observables which are constant in $\Lambda$CDM, or more generally, any Friedman-Lema\^itre-Robertson-Walker (FLRW) model. One of the simplest of these tests is the so-called $O_{\rm m}$ diagnostic, which follows from the Friedman equation in flat $\Lambda$CDM: $O_{\rm m}\equiv [H_0^{-2}H(z)^2-1]/[(1+z)^3-1]= \Omega_{m0}$. Measured deviations of $O_{\rm m}$ from a constant would imply that the flat $\Lambda$CDM assumption is incorrect~\cite{Zunckel:2008ti} (see also \cite{Sahni:2008xx}).  In terms of implementing the null tests, typically the observable~-- in this example the Hubble rate~-- is reconstructed in a model-independent way, using, for example, Gaussian Processes~\cite{Seikel:2012uu,Shafieloo:2012ht}. There are many  other such tests which probe different aspects of $\Lambda$CDM or general FLRW, including the growth of structure and the Cosmological Principle itself (see e.g. \citep{Clarkson:2007pz,Shafieloo:2009hi,Nesseris:2010ep,Seikel:2012cs,Yahya:2013xma,Shafieloo:2018gin,Ghosh:2018ijm,Franco:2019wbj,Bengaly:2020neu,Euclid:2021frk,Bengaly:2021wgc,Dinda:2024kjf,LHuillier:2024rmp,Castello:2024lhl,Dias:2024tpf,Marra:2017pst,Gao:2025ozb,Chiba:2007vm,vonMarttens:2018bvz}).

In this study, we gather together a suite of these tests and reformulate them to be suitable for testing the standard model using DESI BAO (Baryon Acoustic Oscillations) data, together with SNIa and CMB data. This work is an extension of the previous work of two of the authors \citep{Dinda:2024ktd} -- in which deviation from a flat $\Lambda$CDM model was studied mainly through the effective equation of state of dark energy. Here we consider additional diagnostics and include the updated data from DESI DR2 BAO.

In \autoref{sec-flrw}, we consider various identities in the standard flat $\Lambda$CDM model and also in more general FLRW models. Using these identities, we define diagnostics, with their advantages and disadvantages, in \autoref{sec-diagnostics}. 
In \autoref{sec-data}, we briefly discuss the observational data that we use in this analysis and the related observables. Employing improved diagnostics, in \autoref{sec-result} we find the deviations from the standard model corresponding to different combinations of data. These are derived using the reconstructed functions of different observables, which are shown in \autoref{sec-uncertainty}. Finally,  \autoref{sec-conclusion} presents our conclusions. We also briefly discuss kernel dependence of the reconstructed functions in \autoref{sec-kernel}.

\section{{Late-time cosmology}}
\label{sec-flrw}

The goal of this section is to derive identities from the Friedmann equations.
The late-time expansion rate in FLRW models (neglecting radiation) is
\begin{align}
\frac{H^2(z)}{H_0^2} &= \Omega_{\rm m0}(1+z)^3+\Omega_{K0}(1+z)^2+(1-\Omega_{\rm m0}-\Omega_{K0})f_{\rm de}(z) \,,
\label{eq:gnrlH}\\
{f_{\rm de}(z)} &= {\frac{\rho_{\rm de}(z)}{\rho_{\rm de}(0)}}={\exp\left[3\int_{0}^{z}{\rm d}\tilde z\, \frac{1+w(\tilde z)}{1+\tilde z}\right]} \,,
\end{align}
{where $w=p_{\rm de}/\rho_{\rm de}$.}
In $\Lambda$CDM, $f_{\rm de}=1$.

\subsection{Identities for a flat $\Lambda$CDM background}
\label{sec-flatLcdm}

In the flat $\Lambda$CDM model, $\Omega_{K0}=0$ and $f_{\rm de}=1$ in \eqref{eq:gnrlH}. If we know $\Omega_{\rm m0}$ and $H_0$ independently, we know the Hubble time evolution. Alternatively, if we know the Hubble evolution and one of the two parameters independently, we can find the value of the other parameter. In addition to that, we can define many combinations of two mutually independent parameters derived from $\Omega_{\rm m0}$ and $H_0$. For example:
\begin{eqnarray}
\Omega_{\rm m0} &=& \frac{H_0^{-2}{H^2(z)}-1}{(1+z)^3-1} ,
\label{eq:Omegam0} \\
\alpha &\equiv& \Omega_{\rm m0} H_0^2 = \frac{H^2(z)-H_0^2}{(1+z)^3-1} ,
\label{alpha}  \\
H_0^2 &=& H^2(z) - \alpha \left[ (1+z)^3-1 \right] ,
\label{eq:H0sqr} \\
\alpha_2 &\equiv& \frac{1-\Omega_{\rm m0}}{\Omega_{\rm m0}}= \frac{H^2(z)}{\alpha} - (1+z)^3 ,
\label{eq:alpha2} \\
\alpha_3 &\equiv& \left( 1-\Omega_{\rm m0} \right) H_0^2= H^2(z) - \alpha (1+z)^3 \,.
\label{eq:alpha3}
\end{eqnarray}
All these equations are different forms of \eqref{eq:gnrlH} with $\Omega_{K0}=0$ and $f_{\rm de}=1$.

On the other hand, if we know $H(z)$ with the parameters $\Omega_{\rm m0}$ and $H_0$ (or any two mutually independent parameters such as $\alpha$ and $H_0$) independently, we can test an identity in the standard model. From now on, we only consider $\alpha$ and $H_0^2$ among the possible pairs of mutually independent parameters. The first reason is to avoid many combinations and equations. The second and more important reason is that the combination of $\alpha$ and $H_0$ is closer to cosmological observations. For example, $\alpha$ can be computed from early-time observations such as the CMB (cosmic microwave background), independently of $H_0$. On the other hand, $H_0$ can be obtained from local distance ladder observations independent of $\alpha$. Two examples of constrained identities are 
\begin{eqnarray}
&& \frac{H^2(z)-H_0^2}{\alpha[(1+z)^3-1]} = 1 ,
\label{eq:idnt1} \\
&& \frac{ H^2(z) - \alpha (1+z)^3 }{ H_0^2 - \alpha } = 1.
\label{eq:fde_flcdm}
\end{eqnarray}
These are nothing more than \eqref{eq:H0sqr} written differently.

\autoref{eq:gnrlH} is basically the first Friedmann equation. Further identities may be derived from the second Friedmann equation,
\begin{equation}
2H(z)H'(z)  = 3 \alpha (1+z)^2 ,
\label{eq:second_Friedmann}
\end{equation}
where prime is  ${\rm d}/{\rm d} z$. Equation \eqref{eq:second_Friedmann} is equivalent to the derivative of \eqref{eq:alpha3} -- and it is independent of $H_0$. This is simply because we are using extra information from $H'$ and ultimately the degrees of freedom remain the same. Similarly using \eqref{eq:second_Friedmann} and \eqref{eq:H0sqr}, we have
\begin{equation}
H_0^2 = H^2(z) + \frac{2H(z)H'(z) \left[ 1-(1+z)^3
 \right] }{3(1+z)^2} .
\label{eq:identity_deriv_2}
\end{equation}
The above two equations tell us that if we know $H$ and $H'$, we can get $\alpha$ and $H_0$ independently (or any other parameters derived from $\Omega_{\rm m0}$ and $H_0^2$).

If we know either $\alpha$ or $H_0$ alongside $H$ and $H'$ independently, we can find different constrained equations. For example, if we know $H_0$, $H$ and $H'$ independently, we have a constrained equation,
\begin{equation}
\frac{2H(z)H'(z)\left[(1+z)^3-1\right]}{3\left[H^2(z)-H_0^2\right](1+z)^2} = 1 .
\label{eq:identity_deriv_constr_1}
\end{equation}
Similarly, if we know $\alpha$, $H$ and $H'$ independently, another constrained equation is
\begin{equation}
\frac{2H(z)H'(z)}{3\alpha(1+z)^2} = 1 .
\label{eq:identity_deriv_constr_2}
\end{equation}
Further constrained equations which are entirely independent of parameters like $\alpha$ and $H_0$ may be found if we use the extra information from the second Hubble derivative. For example, 
\begin{equation}
\frac{(1+z)\left[H(z)H''(z)+H'{}^2(z)\right]}{2H(z)H'(z)} = 1 .
\label{eq:identity_deriv_constr_3}
\end{equation}

We can derive identities involving integration, related to cosmic distances like comoving distance. In the standard model, the comoving distance is \citep{Dinda:2024kjf,Dinda:2021ffa}
\begin{eqnarray}
D_M (z) &\equiv& c \int_0^z \frac{{\rm d}y}{H(y)} = \frac{c \big[ (1+z)F(z)-F(0) \big] }{H_0 \sqrt{1-\Omega _{\text{m0}}}} ,
\label{eq:DM_LCDM_norad} \\
 \text{where} \quad F(z) &=& \, _2F_1\! \left[ \frac{1}{3},\frac{1}{2};\frac{4}{3};- \frac{\Omega_{\rm m0}(1+z)^3}{1-\Omega_{\rm m0}} \right] ,
\end{eqnarray}
is a hypergeometric function. Using \eqref{eq:DM_LCDM_norad}  we can find many identities in the standard model, such as
\begin{eqnarray}
 \frac{F_{\rm AP}(z) G(z) + H(z) \, _2F_1\!\left[\frac{1}{3},\frac{1}{2};\frac{4}{3};-{\alpha/G^2(z) } \right] }{(1+z) H(z) \, _2F_1\!\left[\frac{1}{3},\frac{1}{2};\frac{4}{3};
-{\alpha (1+z)^3 }/G^2(z)
\right]} = 1 ,
\label{eq:idntt_best} 
\end{eqnarray}
where the Alcock-Paczynski factor is
\begin{eqnarray} 
 F_{\rm AP}(z) = \frac{1}{c}\,H(z) D_M(z) ,
\end{eqnarray}
and
\begin{align} \label{g}
    G(z)=\left[ H^2(z)-\alpha (1+z)^3 \right]^{1/2}\,.
\end{align}
Here we used \eqref{eq:alpha2} and \eqref{eq:alpha3} in \eqref{eq:DM_LCDM_norad}.

\subsection{Identities for perturbations of flat $\Lambda$CDM}
\label{sec-prtbnlcdm}

In the standard model, the growth of matter perturbations at late times obeys
\begin{equation}
\ddot{\delta}_{\rm m} + 2H\dot{\delta}_{\rm m} - 4\pi G \bar{\rho}_{\rm m} \delta_{\rm m} = 0 ,
\label{eq:matter_density_contrast}
\end{equation}
where $\bar{\rho}_{\rm m}$ is the background matter energy density, $\delta_{\rm m}$ is the matter density contrast, which is proportional to the growth function $D_{+}$, and overdots are time derivatives. Then we can rewrite \eqref{eq:matter_density_contrast} as
\begin{equation}
(1+z)^2 D_{+}''(z) + (1+z) A(z) D_{+}'(z) + B(z) D_{+}(z) = 0 ,
\label{eq:grwth_eqn_1}
\end{equation}
where 
\begin{eqnarray}
A(z) &=& (1+z) \frac{H'(z)}{H(z)}-1 = \frac{(1+z)^3-2 \alpha _2}{2 \left[\alpha _2+(1+z)^3\right]} ,
\label{eq:Alcdm} \\
B(z) &=& -\frac{3}{2} \Omega_{\rm m} (z) = -\frac{3 (1+z)^3}{2 \left[\alpha _2+(1+z)^3\right]} .
\label{eq:Blcdm}
\end{eqnarray}
The growing mode solution is 
\begin{equation}
D_{+}(z) = (1+z)^{-1} \left({\, _2F_1\!\left[\frac{1}{3},1;\frac{11}{6};-{\alpha _2}\right]}\right)^{-1} {\, _2F_1\!\left[\frac{1}{3},1;\frac{11}{6};-{\alpha _2}(1+z)^{-3}\right]} ,
\label{eq:D_soln}
\end{equation}
normalised to $D_{+}(0)=1$. The growth rate is
\begin{equation}
f(z) = -\dfrac{{\rm d}\ln{D_{+}}}{{\rm d}\ln{(1+z)}} = 1-\frac{6 \alpha _2}{11 (1+z)^3} \, \frac{_2F_1\!\left[\frac{4}{3},2;\frac{17}{6};-{\alpha _2}(1+z)^{-3}\right]}{ \, _2F_1\!\left[\frac{1}{3},1;\frac{11}{6};-{\alpha _2}(1+z)^{-3}\right]} .
\label{eq:soln_f}
\end{equation}
Using \eqref{eq:alpha2} and \eqref{eq:second_Friedmann} we find a further identity
\begin{eqnarray}
f(z)-1&=&-\frac{ 6 J(z)}{11} \, \frac{_2F_1\left[\frac{4}{3},2;\frac{17}{6};-J(z)\right]}{_2F_1\left[\frac{1}{3},1;\frac{11}{6};-J(z)\right]},
\label{eq:identity_in_f}\\
\label{j}
\text{where} \quad
J(z) &=&\frac{3 H(z)-2 (1+z) H'(z)}{2(1+z)H'(z)}.
\end{eqnarray}

By \autoref{eq:D_soln} and \autoref{eq:soln_f}, using also the relation $\sigma_8(z)=\sigma_{8,0} {D_+(z)}/{D_+(0)}$, we find\footnote{$\sigma_8$ is the power on a scale $R=8h^{-1}\,$Mpc which introduces complications via the dependence on $H_0$, and it is better to use a fixed scale, e.g. $R=12\,$Mpc \cite{Sanchez:2020vvb,Forconi:2025cwp}. We leave this for future work.}
\begin{equation}
f(z)\sigma_8(z) = \sigma_{\rm 8,0} \frac{11 (1+z)^3 \, _2F_1\!\left[\frac{1}{3},1;\frac{11}{6};-{\alpha _2}(1+z)^{-3}\right]-{6\alpha _2} \, _2F_1\!\left[\frac{4}{3},2;\frac{17}{6};-{\alpha _2}(1+z)^{-3}\right]}{11(1+z)^4 \, _2F_1\!\left[\frac{1}{3},1;\frac{11}{6};-\alpha _2\right] } .
\label{eq:fsigma8_soln}
\end{equation}
The derivative gives
\begin{equation}
\frac{3 \, _2F_1\left[\frac{1}{3},2;\frac{11}{6};-\frac{3 H-2 (1+z) H'}{2(1+z)H'}\right] \left[S (1+z) H'+H \left(z S'+S'-2 S\right)\right]}{\, _2F_1\left[\frac{1}{3},1;\frac{11}{6};-\frac{3 H-2 (1+z) H'}{2(1+z)H'}\right] \left[S (1+z) H'+2 H \left(z S'+S'-2 S\right)\right]} = 1 ,
\label{eq:fsigma8_deriv}
\end{equation}
where $S=f\sigma_8$ and we omitted the redshift dependence for brevity.

\subsection{Identities for curved $\Lambda$CDM and FLRW backgrounds}
\label{sec-nonflatLcdm}

In the general case that includes the possibility of spatial curvature, the $\Lambda$CDM evolution is given by $f_{\rm de}=1$ in \eqref{eq:gnrlH} ($\Omega_{\rm K0}\neq0$).
In order to isolate $\Omega_{K0}$, we take a derivative \cite{Seikel:2012cs,Cai:2015pia,Wu:2022fmr,Arjona:2021hmg}:
\begin{equation}
\Omega_{K0} = \frac{-2 z [z (z+3)+3] E(z) E'(z)+3 (z+1)^2 E(z)^2-3 (z+1)^2}{z^2 (z+1) (z+3)}
\,,
\label{eq:Ok0_nonflat_lcdm}
\end{equation}
where $E=H/H_0$. A further derivative gives
\begin{eqnarray}
&& z E(z) \left[z (z+1) (z+3) E''(z)-2 (2 z (z+3)+3) E'(z)\right] \nonumber\\
&& + (z+1) \left[z^2 (z+3) E'(z)^2-3 (z+1)\right]+3 (z+1)^2 E(z)^2 = 0 .
\label{eq:dOk0dz_nonflat_lcdm}
\end{eqnarray}

For a general FLRW model ($\Omega_{\rm K0}\neq0$ and $f_{\rm de}\neq1$), the curvature parameter can be isolated as:
\begin{equation}
\frac{\Omega_{K0}H_0^2}{c^2} = \frac{D^{\prime\,2}_M(z)-D_H^2(z)}{D_H^2(z) D_M^2(z)}.
\label{eq:idnt_nonflat_1}
\end{equation}
A derivative eliminates the constant on the left, leading to an identity that is independent of the value of $\Omega_{K0}$:
\begin{equation}
D_H(z) D_M(z) D_M''(z)+D_H^3(z)-D_H(z) D^{\prime\,2}_M(z)-D_M(z) D_H'(z) D_M'(z) = 0 .
\label{eq:nonflat_flrw_2}
\end{equation}

\section{Diagnostics for null tests}
\label{sec-diagnostics}

\subsection{Null tests of flat $\Lambda$CDM
background}
\label{sec-flatLCDM_diagnostics}

We can consider any of the identities in \autoref{sec-flatLcdm} and test their validity from any relevant combination of observed data. Violation of an identity corresponds to the deviation from the standard model -- this is the concept of a null test. We can have many possible null tests.

\subsubsection*{The popular $O_{\rm m}$ diagnostic}

From \eqref{eq:Omegam0}, we define the diagnostic 
\begin{equation}
O_{\rm m} (z) \equiv \frac{{H_0^{-2}H^2(z)}-1}{(1+z)^3-1} .
\label{eq:Om_diagnostic}
\end{equation}
The null test is that in the standard model, $O_{\rm m} (z)= \Omega_{\rm m0}$.
However, it has several disadvantages:
\begin{itemize}
	\item \textit{Disadvantage I}: It is difficult to quantify whether an evolving parameter is constant or not where the value of that constant is not known a priori.  This is the main disadvantage in this context.
	\item \textit{Disadvantage II}: Most cosmological observations provide data which are directly related to $H$, not $H/H_0$, so that explicit $H_0$ dependence is present and we need prior knowledge of $H_0$. We can use  $H_0$ from observations like the local distance ladder, such as SHOES observations \citep{Riess:2020fzl} or tRGB (tip of the red giant branch) \citep{Freedman:2019jwv}. But there is inconsistency in $H_0$ values from CMB and late-time observations (the `Hubble tension')  and so we should not combine  these observations \citep{Vagnozzi:2019ezj,DiValentino:2021izs,Dinda:2021ffa}. Alternatively we can reconstruct $H_0$ from $H$ using interpolation techniques. However, this will introduce extra errors in the estimate of uncertainty. 
	\item \textit{Disadvantage III}: A minor disadvantage is that the diagnostic is not well defined at $z=0$, although one can consider the limiting value instead of an exact value. 
\end{itemize}

\subsubsection*{Similar diagnostics}
 
We can use  equations \eqref{alpha}--\eqref{eq:alpha3} and \eqref{eq:second_Friedmann} to define similar diagnostics:
\begin{eqnarray}
C_1(z) &=& \frac{H^2(z)-H_0^2}{(1+z)^3-1} ,
\label{eq:diag_C1}  \\
C_2(z) &=& H^2(z) - \alpha \left[ (1+z)^3-1 \right] ,
\label{eq:diag_C2} \\
C_3(z) &=& \frac{H^2(z)}{\alpha} - (1+z)^3 ,
\label{eq:diag_C3} \\
C_4(z) &=& H^2(z) - \alpha (1+z)^3 ,
\label{eq:diag_C4} \\
C_5(z) &=& \frac{2H(z)H'(z)}{3(1+z)^2} ,
\label{eq:diag_C5}
\end{eqnarray}
where $\alpha$ is defined in \eqref{alpha}.
The null tests are $C_i(z)=\,$const, where deviations from a constant value correspond to deviations from the standard model. As with the $O_{\rm m}$ diagnostic, all these diagnostics have \textit{Disadvantage I}.
Only $C_1$ has \textit{Disadvantages II} and \textit{III},  like $O_{\rm m}$.

\begin{itemize}
	\item \textit{Disadvantage IV}: Diagnostics $C_2$, $C_3$, and $C_4$ require the value of $\alpha$. No observations directly give  $\alpha$, nor can we get it from the reconstruction of observables like $H$ or $D_M$, unlike $H_0$. The reason is that $\alpha$ (like $\Omega_{\rm m0}$) is a derived parameter and it requires additional model dependence. For example,  CMB data can indirectly provide its value, depending on an early Universe model. In this sense, diagnostics $C_2$, $C_3$, and $C_4$ are not entirely model-agnostic tests of the standard model. However, they are not very sensitive to any late-time dark energy model.
    \item \textit{Disadvantage V}: The diagnostic $C_5$ is better than the others, since \textit{Disadvantages II, III, IV} do not apply to it. However the main \textit{Disadvantage I} is present -- and it introduces an extra disadvantage, since it depends on $H'$. No cosmological observations directly give $H'$. We need to reconstruct $H$ and then estimate the derivative, introducing extra errors. 
\end{itemize}

\subsubsection*{Other diagnostics}

We can define  diagnostics from \eqref{eq:idnt1} and \eqref{eq:identity_deriv_constr_1}:
\begin{eqnarray}
D_1(z) &=& \frac{H^2(z)-H_0^2}{\alpha[(1+z)^3-1]} ,
\label{eq:D1_diagnostic} \\
D_2(z) &=& \frac{2H(z)H'(z)\left[(1+z)^3-1\right]}{3\left[H(z)^2-H_0^2\right](1+z)^2} .
\label{eq:A2_diagnostic}
\end{eqnarray}
Deviation from $D_i=1$ corresponds to deviation from the standard model. These diagnostics have both \textit{Disadvantage II} ($H_0$ dependence) and \textit{III} (not well defined at $z=0$). Additionally, $D_1$ has \textit{Disadvantage IV} ($\alpha$ dependence) and $D_2$ has \textit{Disadvantage V} ($H'$ dependence). Because the main disadvantage, \textit{Disadvantage I}, is not present, the $D_i$ diagnostics are slightly better than the  $C_i$ diagnostics, but still not good diagnostics because of the $H_0$ dependence.

\subsubsection*{Improved diagnostics}

Using \eqref{eq:fde_flcdm}, \eqref{eq:identity_deriv_constr_2} and \eqref{eq:identity_deriv_constr_3} we can avoid \textit{Disadvantage I}:
\begin{eqnarray}
A_1(z) &=& \frac{ H^2(z) - \alpha (1+z)^3 }{ H_0^2 - \alpha } ,
\label{eq:A1_diagnostic} \\
A_2(z) &=& \frac{2H(z)H'(z)}{3\alpha(1+z)^2} ,
\label{eq:A2_diagnostic} \\
A_3(z) &=& \frac{(1+z)\left[H(z)H''(z)+H'{}^2(z)\right]}{2H(z)H'(z)} .
\label{eq:A3_diagnostic}
\end{eqnarray}
The deviation from $A_i=1$ provides null tests of the standard model. These diagnostics are better since they avoid the main \textit{Disadvantage I} and they are well defined at $z=0$.

\begin{itemize}
    \item \textit{Disadvantage VI}: $A_3$  depends on $H''$, so we have to reconstruct $H$  and perform two derivatives, introducing more errors.
\end{itemize}

The diagnostics up to now do not benefit from cosmological distance data, for example from BAO. Using \eqref{eq:idntt_best}, we define
\begin{equation}
B_1(z) = \frac{F_{\rm AP}(z) G(z) 
+ H(z) \, _2F_1\!\left[\frac{1}{3},\frac{1}{2};\frac{4}{3};-{\alpha/G^2(z) }
\right] }{(1+z) H(z) \, _2F_1\!\left[\frac{1}{3},\frac{1}{2};\frac{4}{3};
-{\alpha (1+z)^3 }/G^2(z)\right]},
\label{eq:B1_diagnostic}
\end{equation}
where $G$ is given by \eqref{g}.
This only has \textit{Disadvantage IV} ($\alpha$ dependence). We can define a similar diagnostic using \eqref{eq:idntt_best} and \eqref{eq:second_Friedmann}:
\begin{eqnarray}
B_2(z) &=& \frac{F_{\text{AP}}(z) K(z) + 3 H(z) \, _2F_1 \left[ \frac{1}{3}, \frac{1}{2}; \frac{4}{3}; -(1+z)^{-3} J^{-1}(z) \right]}{3 (1+z) H(z) \, _2F_1 \left[ \frac{1}{3}, \frac{1}{2}; \frac{4}{3}; -(1+z)^{-3} J^{-1}(z) \right]} ,~~~~~~~
\label{eq:B2_diagnostic} \\
\text{where} \quad
K(z) &=& \big[6(1+z)H(z)H'(z)J(z) \big]^{1/2} ,
\label{K}
\end{eqnarray}
and $J$ is defined by \eqref{j}.
This diagnostic also only has \textit{Disadvantage V} ($H'$ dependence) out of 6 possible disadvantages. Hence $B_1$ and $B_2$ are better than previous diagnostics.

\begin{table}[h!]
\centering
\caption{Summary of background diagnostics and their disadvantages. A good diagnostic mandatorily should not have \textit{Disadvantage I}. In each column corresponding to a disadvantage, a tick sign indicates the presence of that disadvantage.}
\label{table:diagnostics}
\vspace{0.1 cm}
\begin{tabular}{|l|l|}
\hline
\textbf{Disadvantage} & \textbf{Short Description} \\ \hline
 \textit{I} & Difficult to define deviation from  a priori unknown constant \\ \hline
 \textit{II} & Explicit $H_0$ dependence: extra error in uncertainty estimation \\ \hline
 \textit{III} & Not well defined at $z=0$ \\ \hline
 \textit{IV} & $\alpha = \Omega_{\rm m0}H_0^2$ dependence: early-Universe model dependence \\ \hline
 \textit{V} & $H'$ dependence: extra error in uncertainty estimation \\ \hline
 \textit{VI} & $H''$ dependence: extra error in uncertainty estimation \\ \hline
\end{tabular}
\begin{tabular}{|p{2cm}|p{1.5cm}|p{1.5cm}|p{1.5cm}|p{1.5cm}|p{1.5cm}|p{1.5cm}|}
\hline
\textbf{Diagnostic} & \textbf{ I} & \textbf{II} & \textbf{III} & \textbf{IV} & \textbf{V} & \textbf{VI} \\ \hline
$O_{\rm m}$ & \checkmark & \checkmark & \checkmark &  &  &  \\ \hline
$C_1$ & \checkmark & \checkmark & \checkmark &  &  &  \\ \hline
$C_2$ & \checkmark &  &  & \checkmark &  &  \\ \hline
$C_3$ & \checkmark & &  & \checkmark &  &  \\ \hline
$C_4$ & \checkmark &  &  & \checkmark &  &  \\ \hline
$C_5$ & \checkmark &  &  &  & \checkmark &  \\ \hline
\hline
$D_1$ &  & \checkmark & \checkmark & \checkmark &  &  \\ \hline
$D_2$ &  & \checkmark & \checkmark & & \checkmark &  \\ \hline
\hline
$A_1$ &  & \checkmark &  & \checkmark &  &  \\ \hline
$A_2$ &  &  &  & \checkmark & \checkmark &  \\ \hline
$A_3$ &  &  &  &  & \checkmark & \checkmark \\ \hline
\hline
$B_1$ &  &  &  & \checkmark &  &  \\ \hline
$B_2$ &  &  &  &  & \checkmark &  \\ \hline
\end{tabular}
\end{table}

We summarize the 6 disadvantages of background diagnostics in the upper panel of \autoref{table:diagnostics}. In the lower panel, we show the disadvantages of individual diagnostics. 

\subsection{Null tests of flat $\Lambda$CDM
perturbations}
\label{sec-pertn_diagnostics}

If we know $H$, $\alpha$ and $f$, we can a define the diagnostic 
\begin{equation}
G_1(z) = f(z) + \frac{6 G^2(z)}{11 \alpha  (1+z)^3} \, \frac{_2F_1\left[\frac{4}{3},2;\frac{17}{6}; -G^2(z)(1+z)^{-3}/\alpha \right] }{ _2F_1\left[\frac{1}{3},1;\frac{11}{6}; -G^2(z)(1+z)^{-3}/\alpha \right]} .
\label{eq:pG1}
\end{equation}
Equations \eqref{eq:alpha2} and \eqref{eq:soln_f} show that $G_1=1$ in the standard model. Since it is explicitly $\alpha$ dependent, it cannot be computed independently of the early-time cosmological model. Hence it has \textit{Disadvantage IV}. To achieve $\alpha$ independence, we can use
\begin{equation}
G_2(z) = f(z)+\frac{3 K^2(z)}{11 (1+z) H'(z) } \, \frac{_2F_1\left[\frac{4}{3},2;\frac{17}{6}; -J(z) \right]}{ _2F_1\left[\frac{1}{3},1;\frac{11}{6}; -J(z) \right]} ,
\label{eq:pG2}
\end{equation}
where $K$ is defined by \eqref{K}. $G_2=1$ in the standard model by
\eqref{eq:identity_in_f}. Its dependence on $H'$  introduces extra errors and it has \textit{Disadvantage V}.

Using \autoref{eq:fsigma8_deriv}, we define a diagnostic 
\begin{equation}
G_3(z) = \frac{3 \, _2F_1\left[\frac{1}{3},2;\frac{11}{6};-\frac{3 H-2 (1+z) H'}{2(1+z)H'}\right] \left[S (1+z) H'+H \left(z S'+S'-2 S\right)\right]}{\, _2F_1\left[\frac{1}{3},1;\frac{11}{6};-\frac{3 H-2 (1+z) H'}{2(1+z)H'}\right] \left[S (1+z) H'+2 H \left(z S'+S'-2 S\right)\right]} .
\label{eq:pG3}
\end{equation}

\autoref{table:pertn_diagnostics} summarises the disadvantages associated with the perturbation diagnostics.

\begin{table}[h!]
\centering
\caption{Summary of perturbation diagnostics and their disadvantages.}
\label{table:pertn_diagnostics}
\vspace{0.1 cm}
\begin{tabular}{|l|l|}
\hline
\textbf{Disadvantage} & \textbf{Short Description} \\ \hline
\textit{IV} & $\alpha = \Omega_{\rm m0}H_0^2$ dependence: early-Universe model dependence \\ \hline
 \textit{V} & $H'$ dependence: extra error in uncertainty estimation \\ \hline
\end{tabular}
\vspace{0.2 cm}
\begin{tabular}{|p{2cm}|p{1.5cm}|p{1.5cm}|}
\hline
\textbf{Diagnostic} & \textbf{ IV} & \textbf{V} \\ \hline
$G_1$ & \checkmark & \\ \hline
$G_2$ &  & \checkmark \\ \hline
$G_3$ &  & \checkmark \\ \hline
\hline
\end{tabular}
\end{table}

Note that the diagnostics we introduce are designed to test the flat $\Lambda$CDM model within the framework of general relativity and in the linear regime of perturbations. If any of these diagnostics show a deviation, it means the flat $\Lambda$CDM model—as a complete framework—is violated. This model relies on several key assumptions, including general relativity as the theory of gravity, spatial flatness, and a cosmological constant as a form of dark energy, among others. Thus, any deviation could result from a failure in one or more of these assumptions. For example, it could indicate spatial curvature, that dark energy is not a cosmological constant or even a breakdown of general relativity. However, the diagnostics themselves do not reveal which assumption is incorrect—they only indicate that the full flat $\Lambda$CDM model may not hold. This logic applies to all the diagnostics we present, including those based solely on background data (\autoref{table:diagnostics}) and those utilizing both background and perturbation-level data (\autoref{table:pertn_diagnostics}).

\subsection{Null tests of curvature in $\Lambda$CDM and FLRW backgrounds}
\label{sec-curvatureLCDM}

We define a diagnostic corresponding to \eqref{eq:Ok0_nonflat_lcdm}
\begin{equation}
W_{K0} \equiv \Omega_{K0} = \frac{-2 z [z (z+3)+3] E(z) E'(z)+3 (z+1)^2 E(z)^2-3 (z+1)^2}{z^2 (z+1) (z+3)} 
\,.
\label{eq:Wk0}
\end{equation}
Deviation from a constant $W_{K0}$ corresponds to deviation from the curved $\Lambda$CDM. However, the quantification of deviation from a constant value whose a priori value is not known is not a good measure. Therefore we define another diagnostic corresponding to \eqref{eq:dOk0dz_nonflat_lcdm}
\begin{eqnarray}
S_{K0} &=& z E(z) \left[z (z+1) (z+3) E''(z)-2 (2 z (z+3)+3) E'(z)\right] \nonumber\\
&& + (z+1) \left[z^2 (z+3) E'(z)^2-3 (z+1)\right]+3 (z+1)^2 E(z)^2 \,.
\label{eq:Sk0p_dgnst}
\end{eqnarray}
Deviation from $S_{K0}=0$ corresponds to deviation from a general (curved) $\Lambda$CDM.

In a general FLRW model, we can define a diagnostic $S_{K1}$ to test for cosmic curvature:
\begin{equation}
S_{K1} = \frac{D'_M(z)}{D_H(z)} \,.
\label{eq:Sk1_nonflat}
\end{equation}
A deviation from $S_{K1}=1$ corresponds to a deviation from flat FLRW. Interestingly, this is independent of any dark energy model. Similarly, we define corresponding to \eqref{eq:idnt_nonflat_1},
\begin{equation}
W_{K1} \equiv \frac{D^{\prime\,2}_M(z)-D_H^2(z)}{D_H^2(z) D_M^2(z)} \,.
\label{eq:Wk0_nonflat}
\end{equation}
Deviation from $W_{K1}=0$ corresponds to a deviation from flat FLRW. This is also independent of any dark energy model.

Finally, we define another more general diagnostic corresponding to \eqref{eq:nonflat_flrw_2}
\begin{equation}
W_{K2} \equiv D_H(z) D_M(z) D_M''(z)+D_H^3(z)-D_H(z) D^{\prime\,2}_M(z)-D_M(z) D_H'(z) D_M'(z) \,.
\label{eq:nonflat_flrw_test}
\end{equation}
A deviation from $W_{K2}=0$ corresponds to deviation from a general FLRW metric, irrespective of whether curvature is zero or not. This is therefore a test of the Cosmological Principle.

In order to avoid many plots for many diagnostics,  we consider a few diagnostics alongside the $O_m$ diagnostic, to test different cases from actual observations:\\ $A_2$, $B_1$, $B_2$, $G_2$, $G_3$, $W_{\rm K0}$, $S_{\rm K0}$, $S_{\rm K1}$, and $W_{\rm K1}$.

\section{Data and observables}
\label{sec-data}

From DESI DR2 BAO \citep{DESI:2025zgx}, we consider six data points for uncalibrated transverse comoving distance ($\tilde{D}_M$) and six uncalibrated points for radial distance ($\tilde{D}_H$):
\begin{eqnarray}
\tilde{D}_M (z) &=& \frac{D_M(z)}{r_d}= \frac{c}{r_d}\int_0^z \frac{{\rm d}y}{H(y)} , 
\label{eq:BAO_tilde_DM} \\
\tilde{D}_H (z) &=& \frac{D_H(z)}{r_d}=\frac{c}{r_d H(z)} ,
\label{eq:BAO_DH_tilde}
\end{eqnarray}
where $r_d$ is the sound horizon at the baryon drag epoch. The $\tilde{D}_M$ and $\tilde{D}_H$ data are at the same effective redshifts and at each effective redshift they are correlated (see \citep{DESI:2025zgx} for more details). We include these correlations in our analysis.

As an alternative to DESI BAO data, we also consider the BAO measurements from the previous generation spectroscopic surveys of the completed Sloan Digital Sky Survey (SDSS) \citep{eBOSS:2020yzd}.

All the diagnostic variables can be expressed in BAO variables by rewriting $H$, $H'$, $H''$ and $F_{\rm AP}$ as follows:
\begin{eqnarray}
H(z) &=& \frac{c}{r_d \tilde{D}_H(z)} , 
\label{eq:H_BAO} \\
H'(z) &=& -\frac{c \tilde{D}'_H(z)}{r_d \tilde{D}^2_H(z)} , 
\label{eq:Hprime_BAO} \\
H''(z) &=& \frac{c \left[2 \tilde{D}^{\prime\,2}_H(z)-\tilde{D}_H(z) \tilde{D}''_H(z)\right]}{r_d \tilde{D}{}^3_H(z)} , 
\label{eq:Hdblprime_BAO} \\
F_{\rm AP}(z) &=& \frac{\tilde{D}_M(z)}{\tilde{D}_H(z)} .
\label{eq:FAP_BAO}
\end{eqnarray}

We consider the Pantheon+ sample for the apparent magnitude $m_B$ of type Ia supernovae (SNIa), in the redshift range $0.01\leq z\leq2.26$ \citep{Scolnic:2021amr}.  Since we combine BAO and SNIa data in our analysis, we rewrite BAO observables in terms of the SNIa observables:
\begin{eqnarray}
\tilde{D}_M(z) &=& \frac{\beta\,  {\rm e}^{b\, m_B(z)}}{1+z} , 
\label{eq:SN_to_BAO_main} \\
\tilde{D}_H(z) &=& \frac{\beta \, {\rm e}^{b\, m_B(z)} \big[  (1+z)b\, m'_B(z)-1 \big]}{(1+z)^2} , 
\label{eq:SN_to_BAO_DHt} 
\end{eqnarray}
with
\begin{align}
b = \frac{\ln(10)}{5}, \quad \beta = {{\rm e}^{-b(25+M_B)}}\, \frac{1\,\text{Mpc}}{r_d} ,
\label{eq:bao_wrt_SNIa}
\end{align}
where $M_B$ is the absolute peak magnitude of the SNIa. These equations are based on the relation
\begin{equation}
m_B(z) = M_B + 5 \log_{10} \left[ \frac{d_L(z)}{{\rm Mpc}} \right] + 25 ,
\label{eq:mB_vs_dL}
\end{equation}
where $d_L=(1+z)D_M$ is the luminosity distance.

For CMB data we use Planck 2018 TT, TE, EE+lowE+lensing \citep{Planck:2018vyg}, with Atacama Cosmology Telescope (ACT) DR6 data of CMB lensing \citep{ACT:2023kun,ACT:2023dou,Carron:2022eyg}. We do not directly use the full CMB likelihood. Instead, we only use the $\alpha$ and $r_d$ values obtained from the CMB distance priors \citep{Wolf:2024eph,Wolf:2025jlc,Bansal:2025ipo}, for the base $\Lambda$CDM model\footnote{To derive CMB distance priors, we  use the chain {\sf base$_{-}$plikHM$_{-}$TTTEEE$_{-}$lowl$_{-}$lowE$_{-}$lensing} \citep{Zhai:2018vmm}.}. This is because we only need these two parameters in our analysis and they are highly insensitive to late-time dark energy models. The constraints obtained on these parameters are 
\begin{eqnarray}
\alpha &=& 1420.8 \pm 12.2 , 
\label{eq:cmb_alpha} \\
r_d &=& 147.43 \pm 0.25 ~~{{\rm Mpc}} ,
\label{eq:cmb_rd} \\
r\left[\alpha,r_d\right] &=& -0.9 ,
\label{eq:cmb_corr_alpha_rd}
\end{eqnarray}
where $r\left[\alpha,r_d\right]$ is the normalised covariance between $\alpha$ and $r_d$.

For the growth rate, we use 11 uncorrelated $f$ data in the range $0.013 \leq z \leq 1.4$  \citep{Avila:2022xad}.

We  also considered six $f\sigma_8$ data, obtained from SDSS, mentioned in Table III of  \citep{eBOSS:2020yzd}.

Error bars of all these data are 
shown in \autoref{sec-uncertainty}.

\section{Results}
\label{sec-result}

We reconstruct all the relevant observables using Gaussian process regression with zero mean function and squared-exponential kernel covariance function. The details of the methodology can be found in \citep{Dinda:2024ktd}.
Because of the involvement of $H_0$, the reconstruction of $O_{\rm m}$ can be done in two ways.
\begin{itemize}
    \item 
 {We {do not} explicitly use $H_0$ data separately}, e.g from  SHOES \citep{Riess:2020fzl} or tRGB \citep{Freedman:2019jwv}. Then we reconstruct $H$ separately, derived from calibrated BAO data, using \eqref{eq:H_BAO}, or calibrated SNIa data, using \eqref{eq:H_BAO} and \eqref{eq:SN_to_BAO_DHt} -- or both combined. To reconstruct these calibrated distances we need $r_d$ from CMB data, as in \eqref{eq:cmb_rd}. Thus, we need to combine $H_0$ from late-time local observations and $r_d$ from early observations. It is not a good idea to combine these two kinds of observations because of the inconsistencies due to the Hubble tension, which has not been resolved. Thus we avoid these combinations.\\
\item
We do not explicitly use $H_0$ data, but instead reconstruct $H_0$ from reconstructed $H(z)$, i.e., $H_0=H(0)$. Then we do not use any late-time local observations for $H_0$. Consequently, we do not need any early-time observations for $r_d$, because of the ratio ${H}/{H_0}$ in the definition of $O_{\rm m}$  and since we reconstruct $H$ and $H_0$ from the same combination of data. This ratio means that we only need uncalibrated BAO or SNIa data (or both combined, but still uncalibrated) -- as can be seen via ${H}/{H_0}={\tilde{D}_H(0)}/{\tilde{D}_H(z)}$. We can not include CMB data either to reconstruct $O_{\rm m}$ if we do not include local observations of $H_0$. Note that this holds only for a model-agnostic analysis. One can consider a particular model to get $O_{\rm m}(z)$ with the inclusion of CMB data, but here we consider the model-agnostic reconstruction of all the diagnostic variables in order to avoid any bias introduced by the model itself.
\end{itemize}

%%%%%%%%%%%%%%%%%%%%%%%%%%%%%%%%%%
\begin{figure*}
\centering
\includegraphics[height=170pt,width=0.49\textwidth]{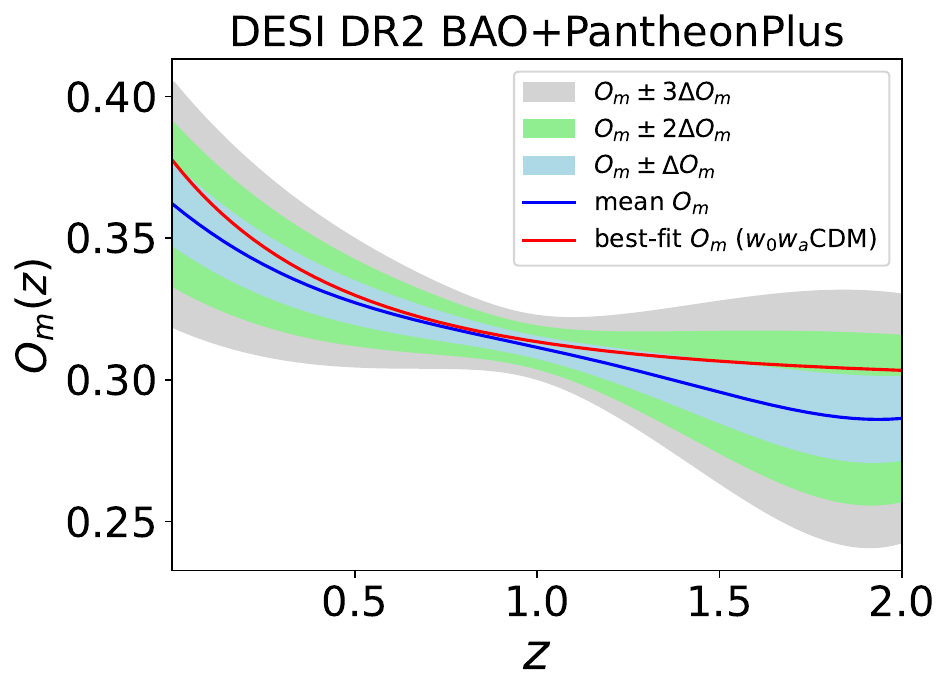}
\includegraphics[height=170pt,width=0.49\textwidth]{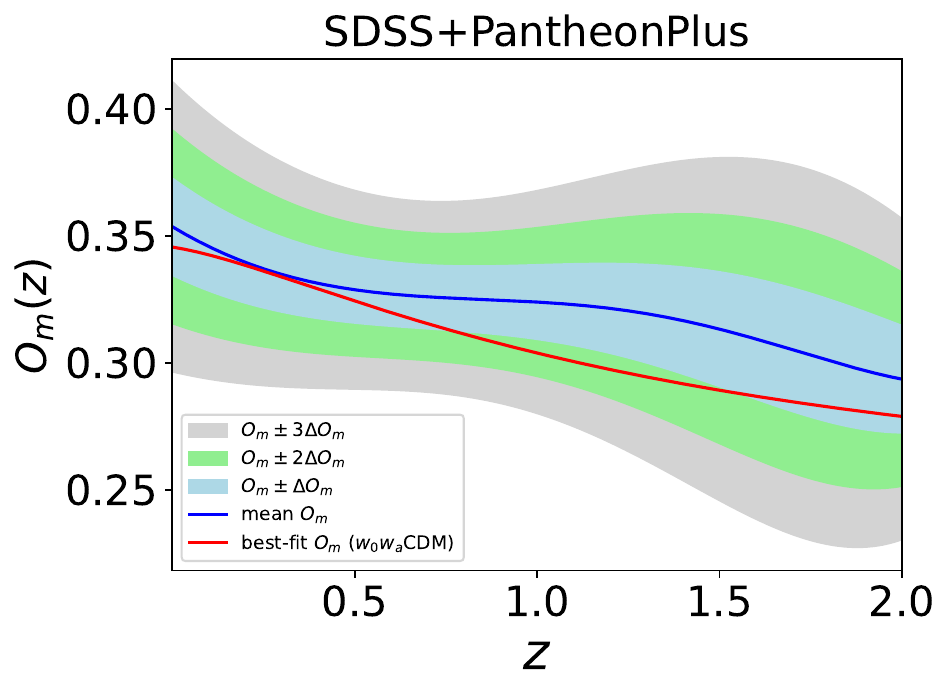}
\caption{
\label{fig:Om}
$O_{\rm m}$ diagnostic from DESI DR2 BAO + PantheonPlus and SDSS + PantheonPlus combinations. Blue lines are reconstructed mean functions of $O_{\rm m}$. Light-blue, green, and grey shading correspond to 1, 2, and 3$\sigma$ confidence regions. Red lines show the best-fit w0waCDM model.
}
\end{figure*}
%%%%%%%%%%%%%%%%%%%%%%%%%%%%%%%%%

\autoref{fig:Om} displays the reconstructed $O_{\rm m}$ diagnostic, corresponding to DESI DR2 BAO + PantheonPlus and SDSS + PantheonPlus combinations of data. The  reconstructed mean of $O_{\rm m}(z)$ (blue lines) and the best-fit $w_0w_a$CDM model (red lines) are also shown. Shaded areas correspond to 1, 2, and 3$\sigma$ confidence intervals. The best-fit values of $(\Omega_{\rm m0},w_0,w_a)$ parameters that we obtain in this model, are $(0.299,-0.888,-0.17)$ and $(0.240,-0.861,-0.30)$ for DESI DR2 BAO + PantheonPlus and SDSS + PantheonPlus  respectively. We see that in both data combinations, there are hints that the reconstructed $O_{\rm m}$ is not a constant. {It is interesting that we find a similar trend in $\Omega_{\rm m0}$ as in previous works such as \citep{Colgain:2024mtg}, even though our result is model-agnostic. We leave this point for future investigation.}

However, we can not quantify the deviation of $O_{\rm m}$ from a constant because we do not know the exact value of the constant $\Omega_{\rm m0}$. This is a consequence of the main \textit{Disadvantage I},  discussed in \autoref{sec-diagnostics}.

%%%%%%%%%%%%%%%%%%%%%%%%%%%%%%%%%%
\begin{figure*}
\centering
\centering
\includegraphics[height=170pt,width=0.49\textwidth]{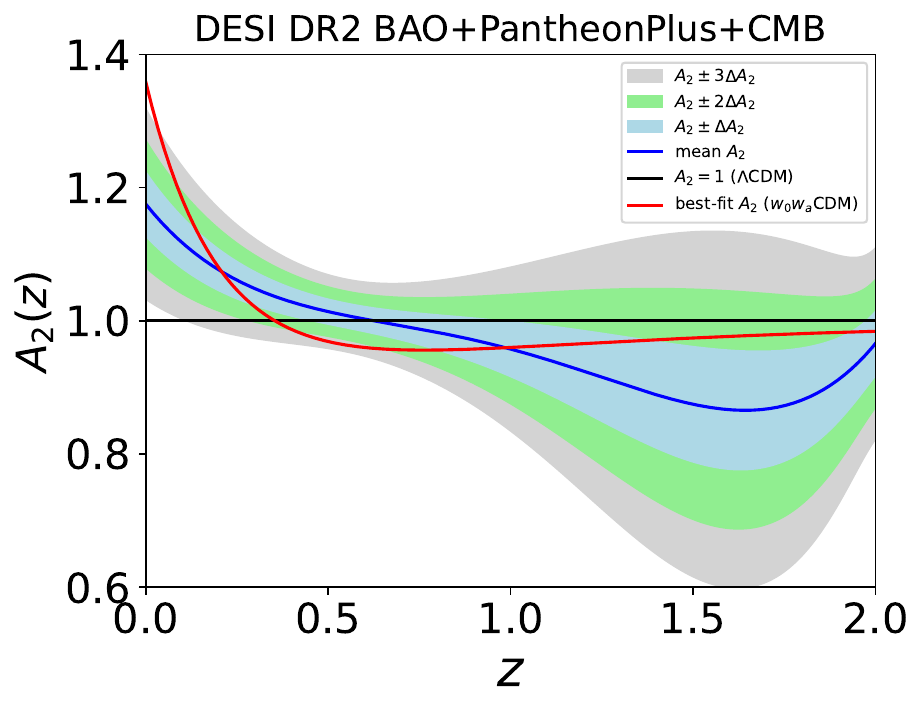}
\includegraphics[height=170pt,width=0.49\textwidth]{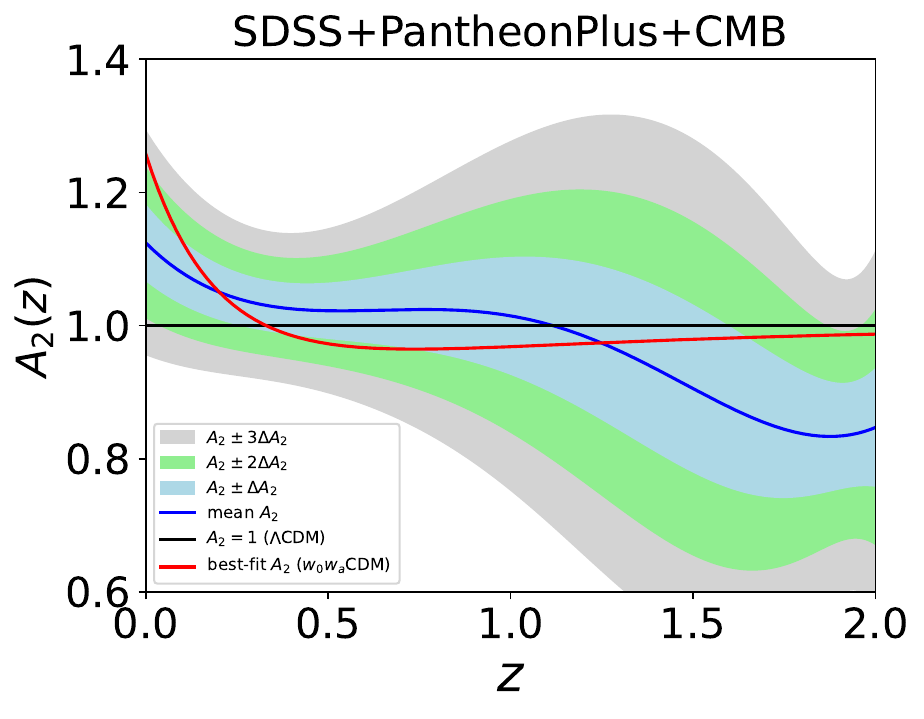}
\caption{
\label{fig:A2}
$A_2$ diagnostic from DESI DR2 BAO + PantheonPlus + CMB and SDSS + PantheonPlus + CMB combinations. Color codes are the same as in \autoref{fig:Om}. The black line shows the standard model value.
}
\end{figure*}
%%%%%%%%%%%%%%%%%%%%%%%%%%%%%%%%%

The same disadvantage is present in diagnostics $C_1,\cdots, C_5$, while diagnostics $D_1, D_2$ suffer from 3 other disadvantages (see \autoref{table:diagnostics}). Hence we focus on $A_2$, $A_3$, $B_1$, and $B_2$.

The reconstructed $A_2$ is presented in \autoref{fig:A2}, using DESI DR2 BAO + PantheonPlus + CMB and SDSS + PantheonPlus + CMB combinations of data.  The best-fit values of $(\Omega_{\rm m0},w_0,w_a)$ parameters are $(0.3114,-0.838,-0.62)$ and $(0.3161,-0.881,-0.48)$ for DESI DR2 BAO + PantheonPlus + CMB and SDSS + PantheonPlus + CMB, respectively. For this diagnostic, we know the exact value, $A_2=1$, for the $\Lambda$CDM model, so we can measure the deviations of this model from the reconstructed mean. At higher redshifts ($z\gtrsim1$), the deviations are around 1 to 1.5$\sigma$ for DESI DR2 BAO + PantheonPlus + CMB. At lower redshifts, deviation increases with decreasing redshift and at $z=0$ it is maximum at more than $3\sigma$. For SDSS + PantheonPlus + CMB, the deviations are $\sim2\sigma$ at redshifts close to $z=0$ and $z=2$, while in between, the deviations are $<1\sigma$. We observe that, at lower redshifts, the red lines show larger deviations compared to the blue lines. This indicates that the deviations are more significant in the flat $w_0w_a$CDM model than in the GP-based predictions. Notably, these deviations in the flat $w_0w_a$CDM model are consistent with the DESI DR2 BAO results for the same model and the same combination of data.

%%%%%%%%%%%%%%%%%%%%%%%%%%%%%%%%%%
\begin{figure*}
\centering
\includegraphics[height=170pt,width=0.49\textwidth]{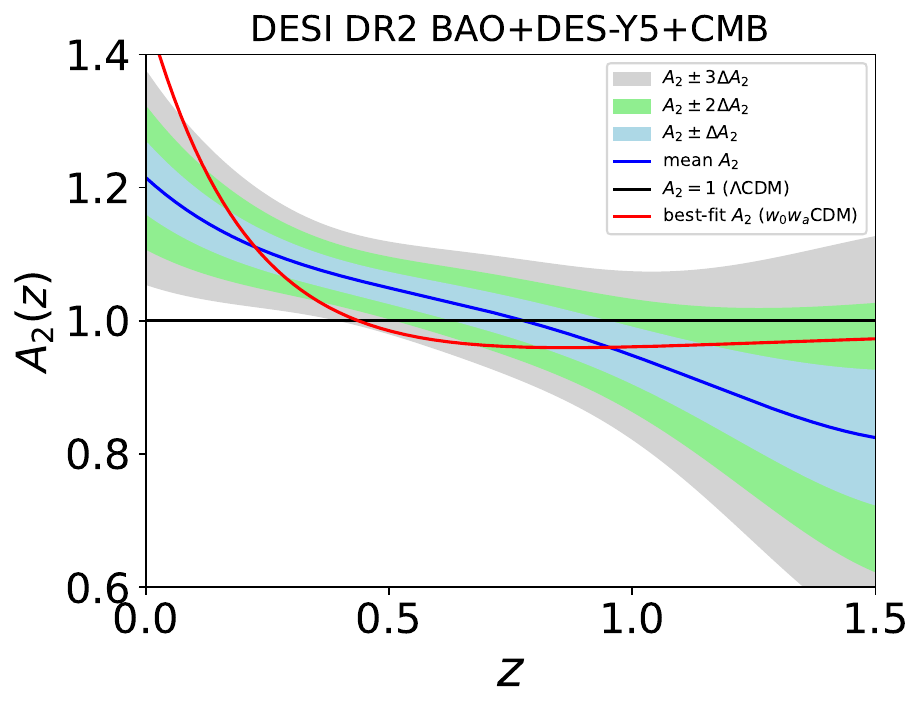}
\caption{
\label{fig:A2_DES5}
$A_2$ diagnostic from DESI DR2 BAO + DES-Y5 + CMB combination of data.
}
\end{figure*}
%%%%%%%%%%%%%%%%%%%%%%%%%%%%%%%%%

To complete the discussion of the $A_2$ results, we also consider the DES-Y5 supernova dataset \citep{DES:2024jxu} as an alternative to PantheonPlus. In \autoref{fig:A2_DES5}, we show the $A_2$ diagnostic for the combination DESI DR2 BAO + DES-Y5 + CMB. Comparing the left panel of \autoref{fig:A2} with \autoref{fig:A2_DES5}, we find that the deviations are similar at higher redshifts. However, at lower redshifts, the deviations are significantly larger for the DESI DR2 BAO + DES-Y5 + CMB case compared to DESI DR2 BAO + PantheonPlus + CMB. For DESI DR2 BAO + DES-Y5 + CMB, the deviation exceeds $3\sigma$ at low redshift, reaching nearly $4\sigma$ near $z = 0$. This behavior is consistent with the trends reported in the DESI DR2 BAO main results.

%%%%%%%%%%%%%%%%%%%%%%%%%%%%%%%%%%
\begin{figure*}
\centering
\centering
\includegraphics[height=170pt,width=0.49\textwidth]{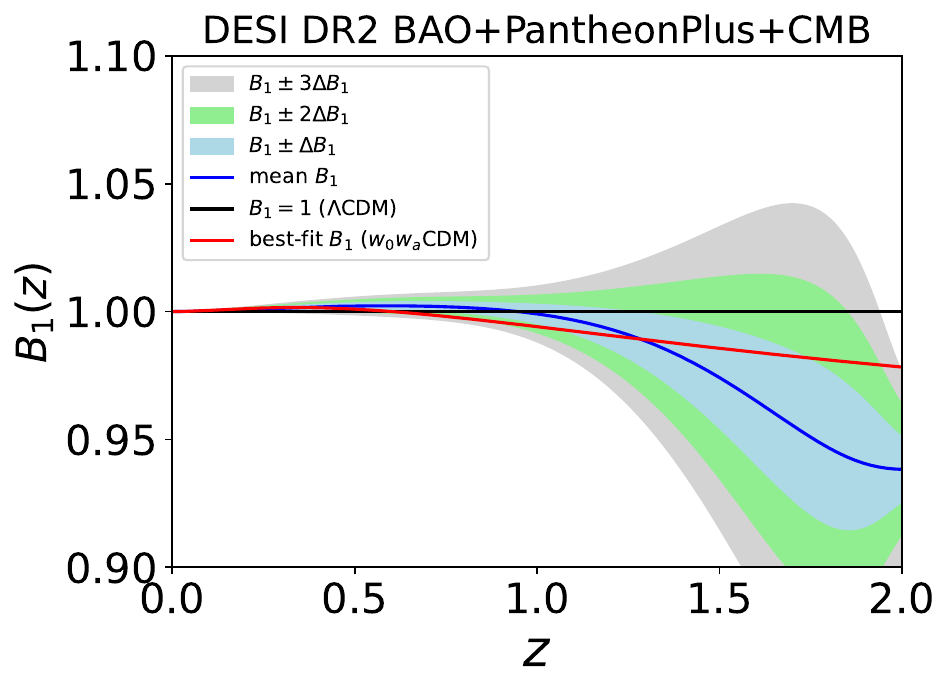}
\includegraphics[height=170pt,width=0.49\textwidth]{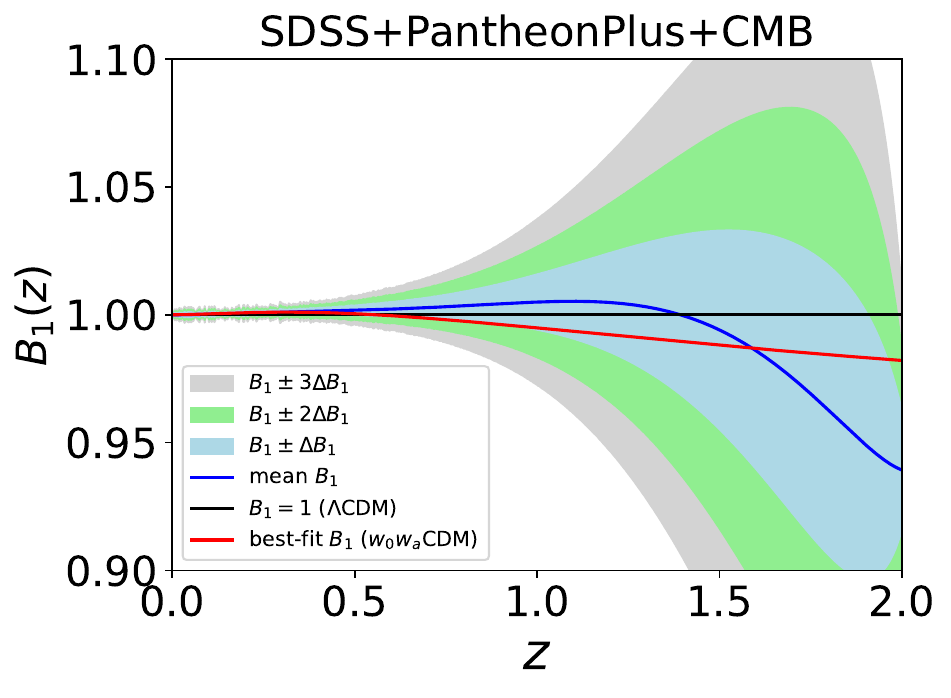}
\caption{
\label{fig:B1}
$B_1$ diagnostic using DESI DR2 BAO + PantheonPlus + CMB and SDSS + PantheonPlus + CMB combinations.
}
\end{figure*}
%%%%%%%%%%%%%%%%%%%%%%%%%%%%%%%%%

%%%%%%%%%%%%%%%%%%%%%%%%%%%%%%%%%%
\begin{figure*}
\centering
\centering
\includegraphics[height=170pt,width=0.49\textwidth]{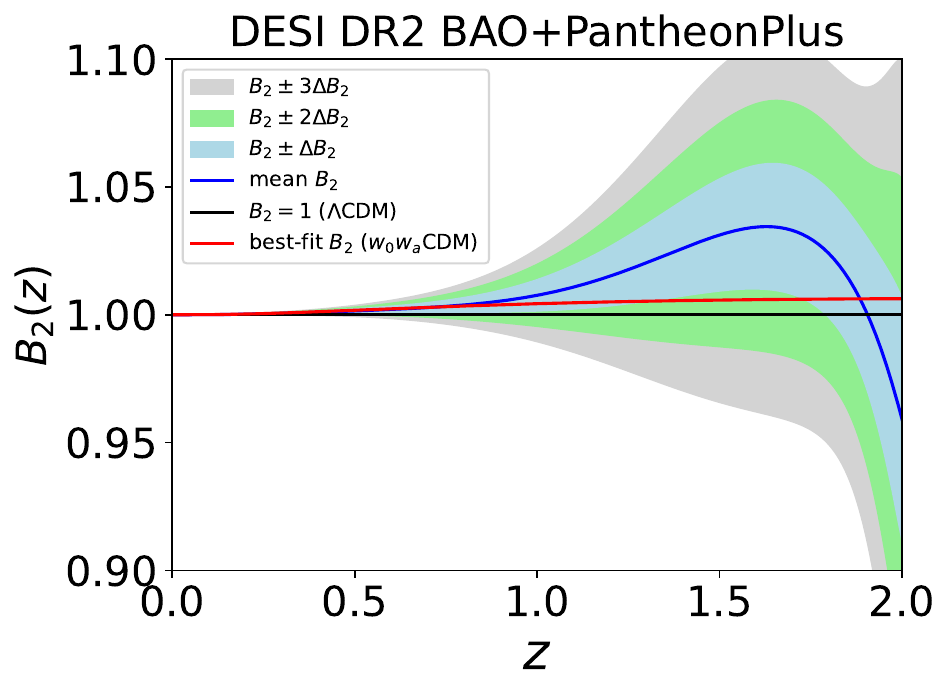}
\includegraphics[height=170pt,width=0.49\textwidth]{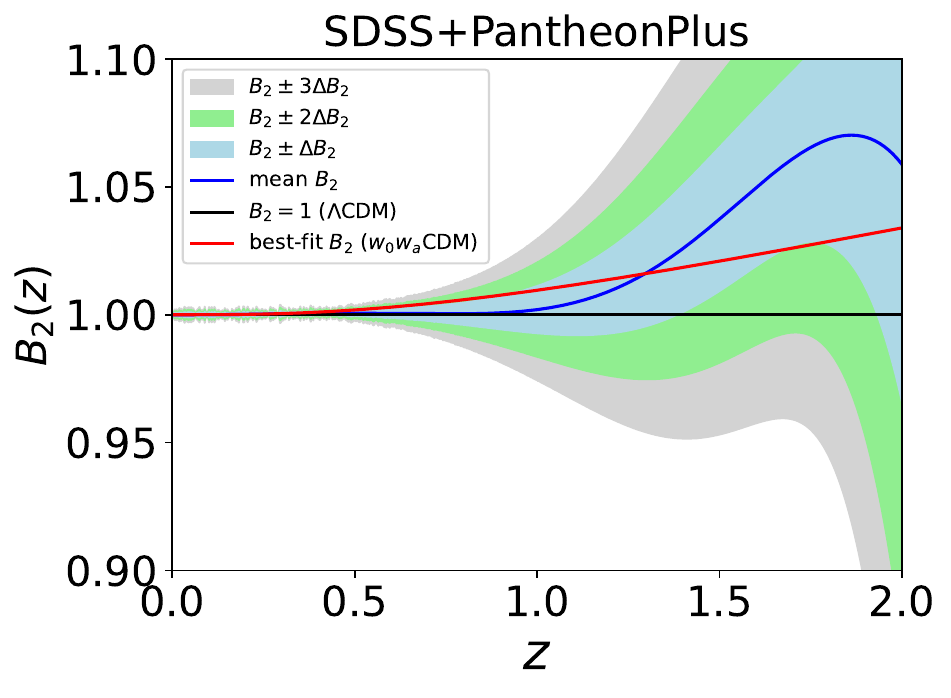}
\caption{
\label{fig:B2}
$B_2$ diagnostic from DESI DR2 BAO + PantheonPlus and SDSS + PantheonPlus combinations.
}
\end{figure*}
%%%%%%%%%%%%%%%%%%%%%%%%%%%%%%%%%

\autoref{fig:B1} and \autoref{fig:B2} display the reconstructed $B_1$ and $B_2$ diagnostics, using DESI DR2 BAO and SDSS BAO data combined with  with PantheonPlus + CMB ($B_1$), or PantheonPlus only ($B_2$). This is because $B_1$ is reconstructed from calibrated BAO (or SNIa or combined) data, whereas $B_2$  is reconstructed from uncalibrated data. It is evident that $B_1$ (like $A_2$) is only late-time model-agnostic while $B_2$ (like $A_3$) is both late-time and early-time model-agnostic. Deviations from the $\Lambda$CDM model are around 1 to 2$\sigma$, except for high redshift (deviations are a little higher) for the DESI DR2 BAO case. The deviations are well within 1$\sigma$ except at higher redshifts (around $z \sim 1.8$), with deviation $\sim2\sigma$ in the $B_2$ SDSS case. Note that the deviations in the $B_1$ and $B_2$ diagnostics are nearly zero at very low redshift. This is because both are related to cosmological distances, which are defined to be zero at $z = 0$. As a result, different cases show no significant differences near $z = 0$. Deviations start to appear gradually as the redshift increases. Also, other quantities contribute to these diagnostics, making the deviations evolve nonlinearly with redshift. These combined effects are reflected in the behavior of the $B_1$ and $B_2$ diagnostics.

%%%%%%%%%%%%%%%%%%%%%%%%%%%%%%%%%%
\begin{figure*}
\centering
\centering
\includegraphics[height=170pt,width=0.49\textwidth]{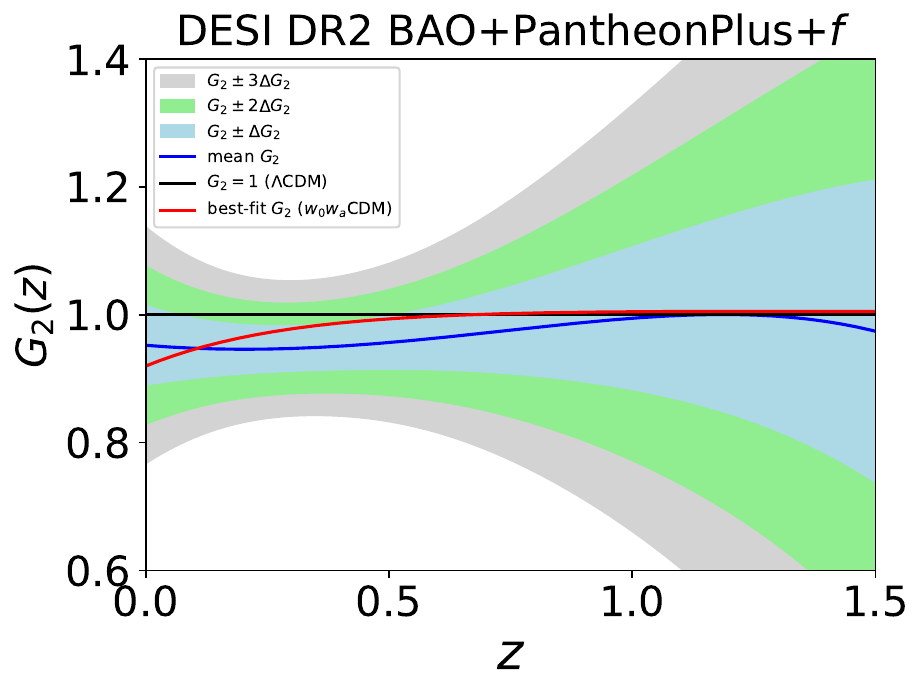}
\includegraphics[height=170pt,width=0.49\textwidth]{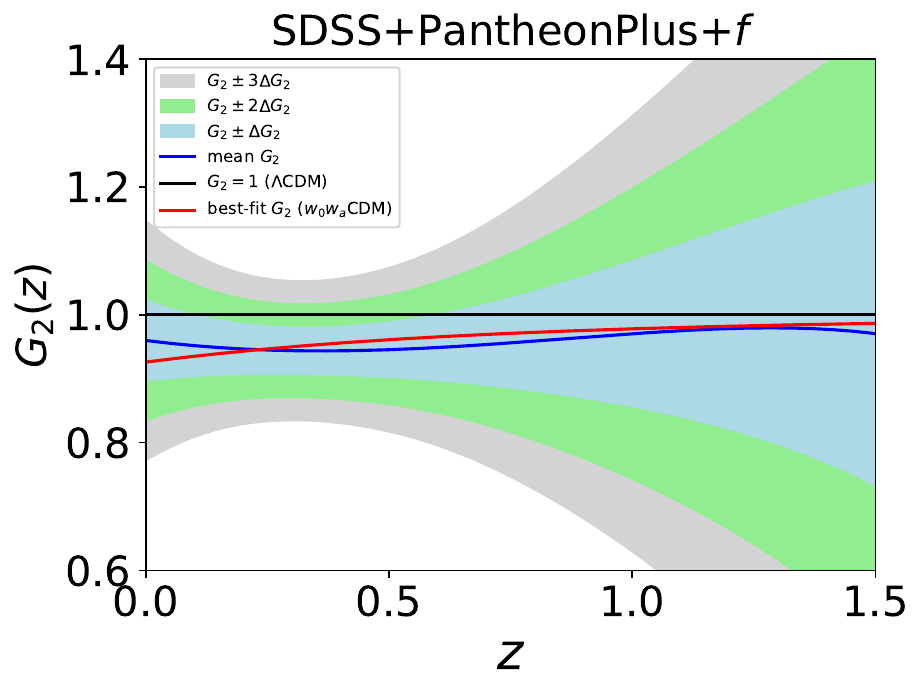}
\caption{
\label{fig:G2}
$G_2$ diagnostic from DESI DR2 BAO + PantheonPlus+$f$ and SDSS + PantheonPlus+$f$ data.
}
\end{figure*}
%%%%%%%%%%%%%%%%%%%%%%%%%%%%%%%%%

We present the perturbation diagnostic $G_2$ in \autoref{fig:G2}, based on the combinations DESI DR2 BAO + PantheonPlus+$f$ and SDSS + PantheonPlus+$f$. For $(\Omega_{\rm m0},w_0,w_a)$, we obtain the best-fit values  $(0.302,-0.875,-0.31)$ and $(0.287,-0.892,0.02)$ for DESI DR2 BAO + PantheonPlus + $f$ and SDSS + PantheonPlus + $f$ respectively. We find that the $\Lambda$CDM model is within $\sim1\sigma$ confidence.

%%%%%%%%%%%%%%%%%%%%%%%%%%%%%%%%%%
\begin{figure*}
\centering
\includegraphics[height=170pt,width=0.49\textwidth]{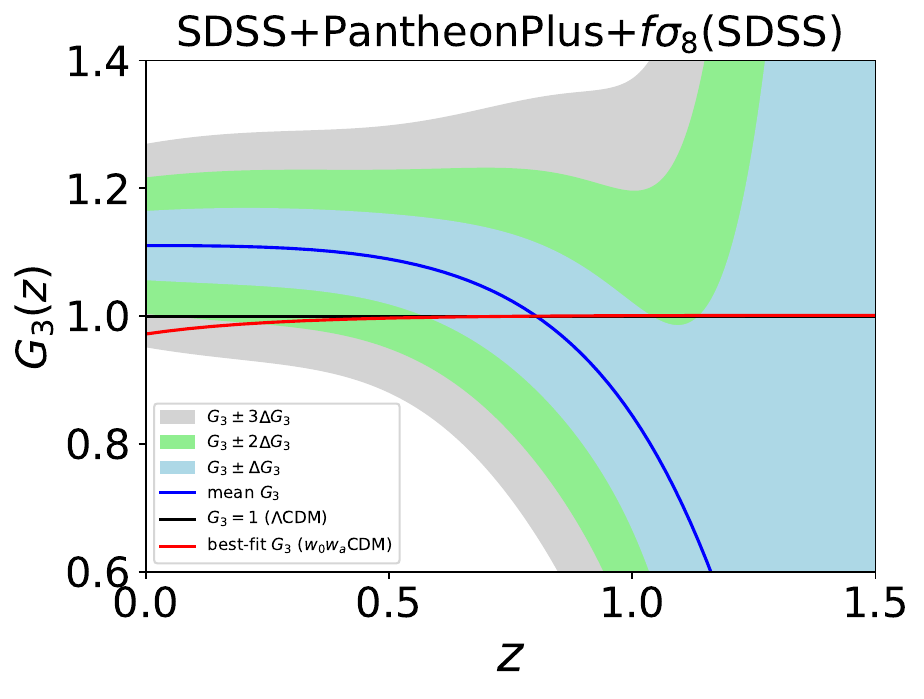}
\caption{
\label{fig:G3}
$G_3$ diagnostic corresponding to SDSS+PantheonPlus+$f\sigma_8$(SDSS) combination of data. Colour codes are the same as in Fig.~\ref{fig:Om}.
}
\end{figure*}
%%%%%%%%%%%%%%%%%%%%%%%%%%%%%%%%%

We plot the $G_3$ diagnostic in Fig.~\ref{fig:G3}, using theSDSS+PantheonPlus+$f\sigma_8$(SDSS) combination of data. The colour codes are the same as in previous figures. The red line is the best-fit flat $w_0w_a$CDM model for the same combination of data. The best-fit values of $(\Omega_{\rm m0},w_0,w_a)$ that we obtain in this model are $(0.260,-0.881,-0.27)$. We find that the $\Lambda$CDM model is  within 1$\sigma$ and 2$\sigma$ confidence regions for $z>0.5$ and $z<0.5$ respectively.

%%%%%%%%%%%%%%%%%%%%%%%%%%%%%%%%%%
\begin{figure*}
\centering
\centering
\includegraphics[height=170pt,width=0.49\textwidth]{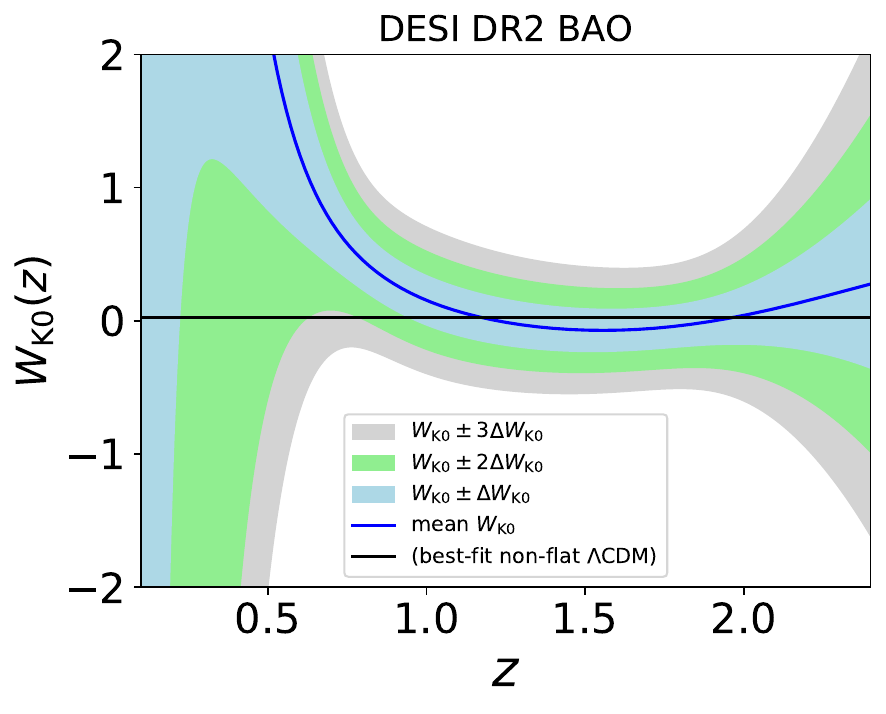}
\includegraphics[height=170pt,width=0.49\textwidth]{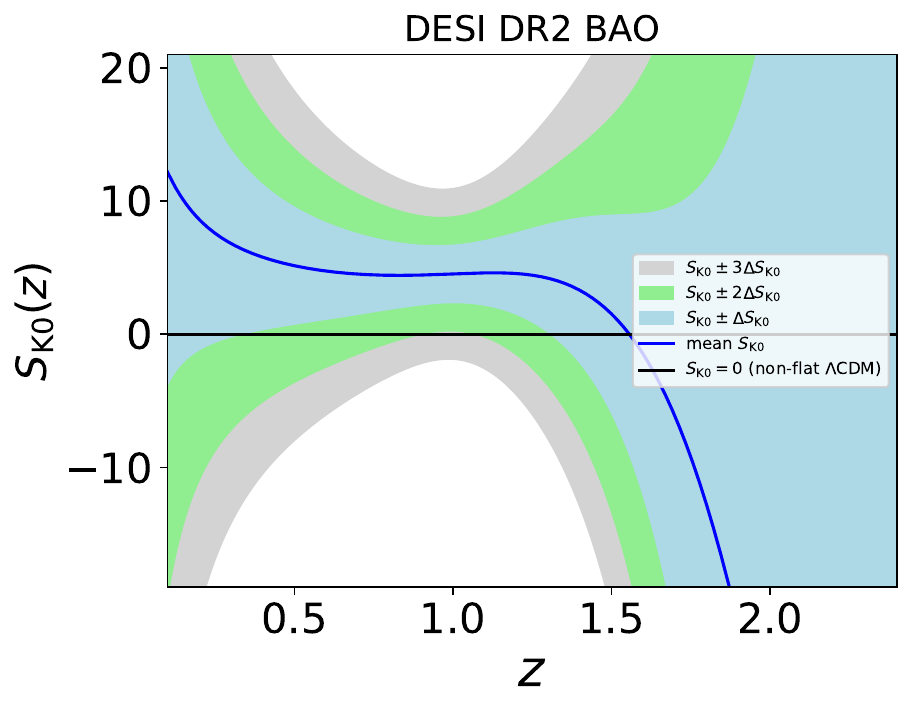}
\caption{
\label{fig:WK0_SK0}
$W_{K0}$ (left) and $S_{K0}$ (right) diagnostics corresponding to DESI DR2 BAO data.
}
\end{figure*}
%%%%%%%%%%%%%%%%%%%%%%%%%%%%%%%%%

Now we briefly test for deviations from a curved $\Lambda$CDM. In \autoref{fig:WK0_SK0}, the left panel shows $W_{K0}$ diagnostics for DESI DR2 BAO data. We can see the deviations from a constant $W_{K0}$ are not significant at higher redshifts but are moderate at lower redshifts. To exactly measure deviation from a curved $\Lambda$CDM, we show deviations from it through the $S_{K0}$ diagnostic, in the right panel. We see the deviation is moderate at lower redshifts, at around 1 to 2$\sigma$.

%%%%%%%%%%%%%%%%%%%%%%%%%%%%%%%%%%
\begin{figure*}
\centering
\centering
\includegraphics[height=170pt,width=0.49\textwidth]{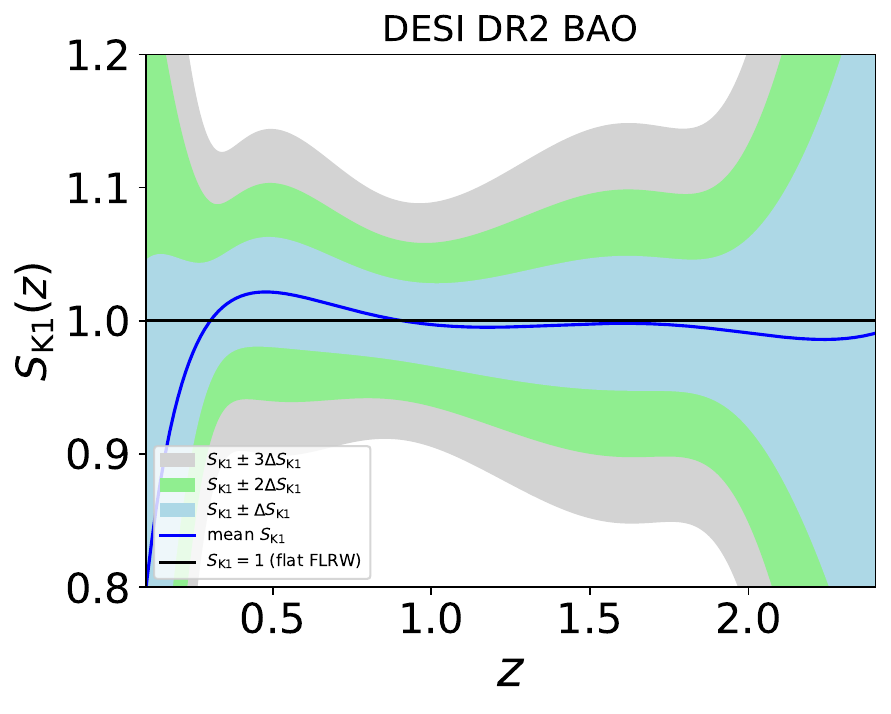}
\includegraphics[height=170pt,width=0.49\textwidth]{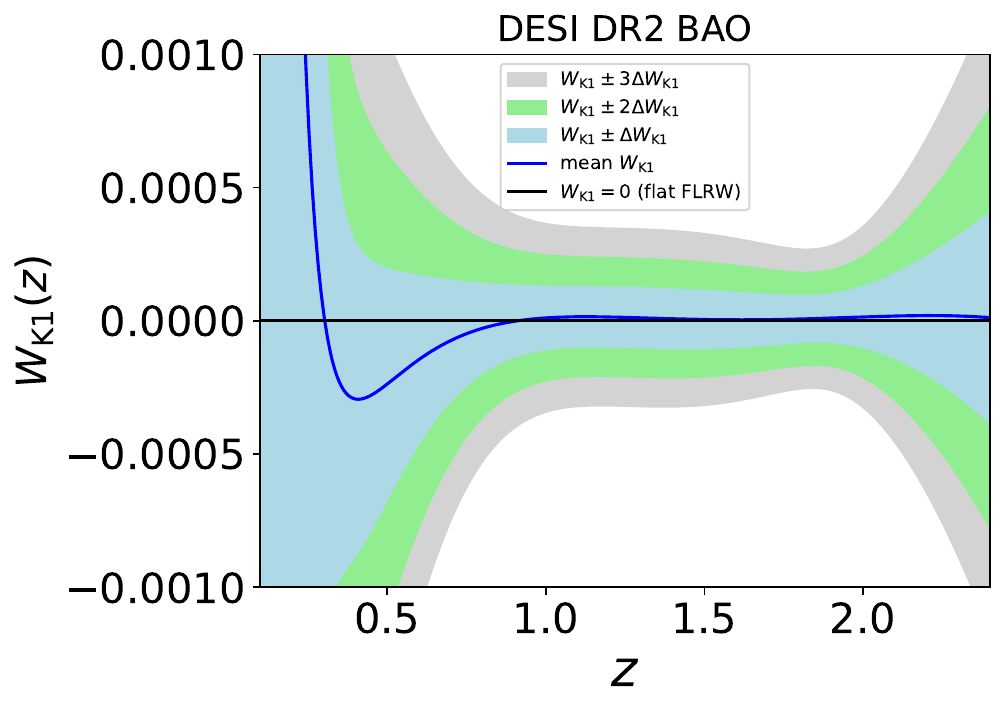}
\caption{
\label{fig:SK1_WK1}
$S_{K1}$ (left) and $W_{K1}$ (right) diagnostics corresponding to DESI DR2 BAO data.
}
\end{figure*}
%%%%%%%%%%%%%%%%%%%%%%%%%%%%%%%%%

Next, we check for deviation from flatness in an FLRW model through two diagnostics $S_{K1}$ (left) and $W_{K1}$ (right) in \autoref{fig:SK1_WK1}. We find there is no evidence for this.

%%%%%%%%%%%%%%%%%%%%%%%%%%%%%%%%%%
\begin{figure*}
\centering
\centering
\includegraphics[height=170pt,width=0.49\textwidth]{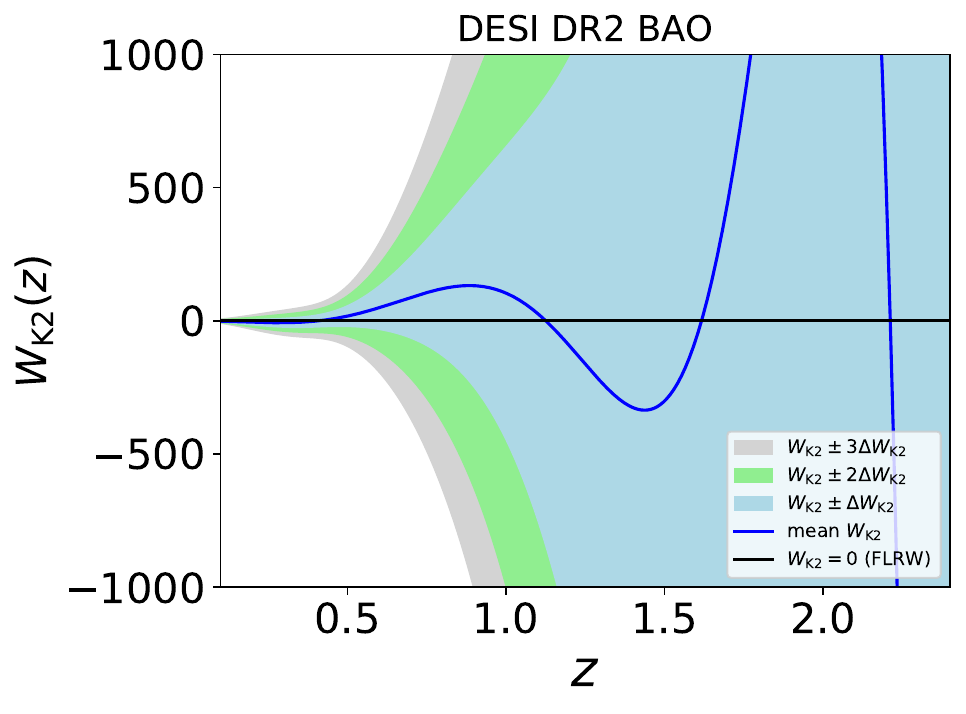}
\includegraphics[height=170pt,width=0.49\textwidth]{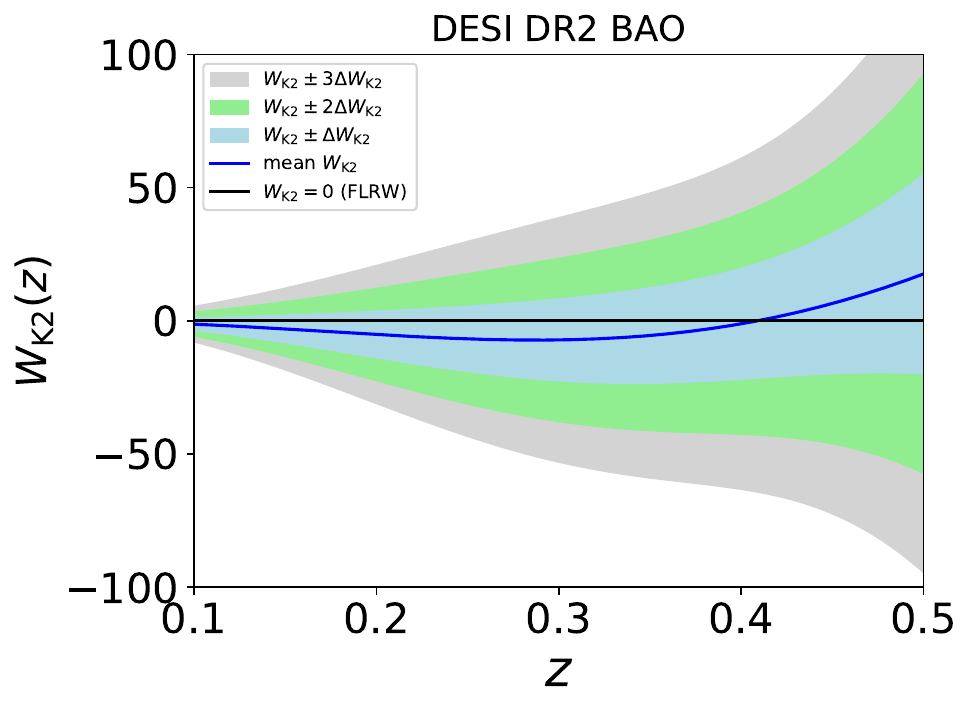}
\caption{
\label{fig:WK2}
$W_{K2}$ diagnostic corresponding to DESI DR2 BAO data.
}
\end{figure*}
%%%%%%%%%%%%%%%%%%%%%%%%%%%%%%%%%

Finally, we test for deviation from a general FLRW metric (irrespective of curvature) through the $W_{K2}$ diagnostic in \autoref{fig:WK2}. The right panel is the zoomed-in version of the left panel and is for lower redshifts only. There is no evidence for deviations from a more general FLRW metric either.

Each diagnostic has a different deviation at different redshifts. There is no hard and fast rule to define an average deviation from all these different deviations. Here  we propose a reasonable definition: the average confidence interval $\left< \sigma \right>$ is the square root of the total chi-squared, normalized to the total number of redshift points $N$:
\begin{equation}
\left< \sigma \right> = \left({\frac{\chi^2}{N}}\right)^{1/2} \quad \text{with} \quad \chi^2 = \bm{v}^{\sf T} \bm{C}^{-1} \bm{v}\,.
\label{eq:avg_sigma}
\end{equation}
Here $\bm v$ is the vector of deviations from $\Lambda$CDM (or a more general model accordingly) and $\bm C$ is the covariance from the uncertainty estimation. \autoref{table:avg_sigma} shows $\left< \sigma \right>$ for different diagnostics and data combinations. We find the average deviations range from 0.27 to 1.76$\sigma$.

%%%%%%%%%%%%%%%%%%%%%%%%%%%%%%%%%%%%%%%%%%%%%%%%%%%%%%%%%%%%%
\begin{table*}
\begin{center}
DESI DR2 BAO + PantheonPlus + CMB \\
\vspace{0.2cm}
\begin{tabular}{|c|c|}
\hline
diagnostic & $\left< \sigma \right>$ \\
\hline
$A_2$ & 1.45 \\
$B_1$ & 1.76 \\
\hline
\end{tabular} \\
\vspace{0.5cm}
DESI DR2 BAO + PantheonPlus \\
\vspace{0.2cm}
\begin{tabular}{|c|c|}
\hline
diagnostic & $\left< \sigma \right>$ \\
\hline
$B_2$ & 1.54 \\
\hline
\end{tabular} \\
\vspace{0.5cm}
DESI DR2 BAO \\
\vspace{0.2cm}
\begin{tabular}{|c|c|}
\hline
diagnostic & $\left< \sigma \right>$ \\
\hline
$W_{\rm K0}$ & 1.02 \\
$S_{\rm K0}$ & 1.12 \\
$S_{\rm K1}$ & 0.30 \\
$W_{\rm K1}$ & 0.27 \\
$W_{\rm K2}$ & 0.31 \\
\hline
\end{tabular}
\end{center}
\caption{
Average $\sigma$ deviation in different diagnostics.
}
\label{table:avg_sigma}
\end{table*}
%%%%%%%%%%%%%%%%%%%%%%%%%%%%%%%%%%%%%%%%%%%%%%%%%%%%%%%%%%%%%%%%%%%

\section{Conclusions}
\label{sec-conclusion}

The deviation from the standard $\Lambda$CDM cosmological model can be studied through null tests, without assuming any model or parametrization, based on diagnostic variables such $O_{\rm m}$. This approach is important to avoid any model-dependent bias.

For the $O_{\rm m}$ diagnostic, the standard model is tested through the reconstructed $O_{\rm m}$, in order to check whether the reconstructed value is constant. Even if we see these values are not constant over a particular redshift interval, we can not measure how much $\Lambda$CDM  is ruled out, because we do not exactly know a priori the value of $\Omega_{\rm m0}$. This is a significant problem in probing deviations from the $\Lambda$CDM model in a data-driven reconstruction. We discussed several other similar diagnostics which have the same problems.

We next discussed how to construct better diagnostics from different identities in the $\Lambda$CDM model. We considered the advantages and disadvantages of several new diagnostic variables -- and also how one can define customized diagnostics for different purposes. We provided a detailed methodology for this.

As examples, we focused on four improved diagnostics of the background cosmology -- $A_2$, $A_3$, $B_1$, and $B_2$ -- and a new perturbation diagnostic $G_2$. One can define many other possible diagnostics combining these diagnostics.

With these improved diagnostics, we studied deviations of the $\Lambda$CDM model from the observed data of DESI DR2 BAO, SDSS BAO, Pantheon+ SNIa, Planck 2018 CMB data (including ACT DR6 CMB lensing data), and growth rate ($f$) data. We found that the deviations are around $1\sigma$ in most of the redshift interval (in a few redshift ranges it is more than 1$\sigma$). These deviations indicate that the evidence for dynamical dark energy is not particularly strong, and is weaker than that inferred from the $w_0w_a$ parametrization.

We also briefly test for curvature in a $\Lambda$CDM model. We found low evidence (less than 2$\sigma$). Furthermore, there is no evidence for curvature in FLRW or for a deviation from the general FLRW model -- i.e. no evidence for a violation of the Cosmological Principle.

\acknowledgments
BRD and RM are supported by the South African Radio Astronomy Observatory and the National Research Foundation (Grant No. 75415).
BRD, RM and SS acknowledge support for this work from the University of Missouri South African Education Program. 
SS acknowledges support for this work from NSF-2219212. 
SS is supported in part by World Premier International Research Center Initiative, MEXT, Japan.

\clearpage

\appendix

\section{Reconstructed functions and error estimations}
\label{sec-uncertainty}

When we have different data of multiple observables, we need to use multi-task Gaussian Process (GP) instead of the standard single-task GP -- especially if the data have correlations among different observables. This is the case for BAO observations where $\tilde{D}_M$ and $\tilde{D}_H$ are correlated. These correlations are very large (of the order of $0.5$ for the normalized covariance) so we cannot neglect them and use single-task GP for each of them separately. Here we briefly discuss multi-task GP. We use a special kind of multi-task GP where one observable ($\tilde{D}_H$) is related to the derivative of another observable ($\tilde{D}_M$). For details of the methodology, we follow  \citep{Dinda:2024ktd}. A brief summary is as follows.

We consider a multi-task GP framework, where the function values $Y_1 = Y(X_1)$ and their first derivatives $Y'_2 = Y'(X_2)$ are modeled jointly. The joint distribution of these quantities is given by a multivariate Gaussian:
\begin{equation}
\tilde{Y}_1 \sim \mathcal{N} \left( \tilde{\mu}_1, \tilde{\Sigma}_{11} \right),
\label{eq:double_norm}
\end{equation}
where
\begin{equation}
\tilde{Y}_1 =
\begin{bmatrix}
Y_1 \\
Y'_2
\end{bmatrix}
=
\begin{bmatrix}
Y(X_1) \\
Y'(X_2)
\end{bmatrix},
\label{eq:tilde_Y}
\end{equation}
\begin{equation}
\tilde{\mu}_1 =
\begin{bmatrix}
\mu_1 \\
\mu'_2
\end{bmatrix}
=
\begin{bmatrix}
\mu(X_1) \\
\mu'(X_2)
\end{bmatrix},
\label{eq:tilde_mu}
\end{equation}
\begin{equation}
\tilde{\Sigma}_{11} =
\begin{bmatrix}
\Sigma_{11} & \Sigma_{12}^{(0,1)} \\
\Sigma_{21}^{(1,0)} & \Sigma_{22}^{(1,1)}
\end{bmatrix}.
\label{eq:tilde_Sigma}
\end{equation}
In equation \eqref{eq:tilde_Sigma}, the covariance matrix blocks are defined as follows:
\begin{align}
\Sigma_{11} &= K_{11} + C_{11} \qquad \qquad \qquad \qquad\text{(function-function)},
\label{eq:Sigma}\\
\Sigma_{12}^{(0,1)} &= K_{12}^{(0,1)} + C_{12}^{(0,1)} \qquad \quad \qquad \quad~  \text{(function-derivative)},
\label{eq:Sigma_p}\\
\Sigma_{21}^{(1,0)} &= K_{21}^{(1,0)} + C_{21}^{(1,0)} = \left(\Sigma_{12}^{(0,1)}\right)^{\mathrm{T}} ~~ \text{(derivative-function)},
\label{eq:Sigma_p_T}\\
\Sigma_{22}^{(1,1)} &= K_{22}^{(1,1)} + C_{22}^{(1,1)} \quad\qquad\qquad\quad~ \text{(derivative-derivative)}.
\label{eq:Sigma_p_p}
\end{align}
Here the $K$ terms represent kernel-based covariance contributions and the $C$ terms account for observational uncertainties. The GP is trained by maximizing the likelihood of the joint observations, which is equivalent to minimizing the negative log-marginal likelihood:
\begin{equation}
- \log P(\tilde{Y}_1|\tilde{X}_1) = \frac{1}{2} (\tilde{Y}_1 - \tilde{\mu}_1)^{\mathrm{T}} \tilde{\Sigma}_{11}^{-1} (\tilde{Y}_1 - \tilde{\mu}_1)
+ \frac{1}{2} \log |\tilde{\Sigma}_{11}|
+ \frac{\tilde{n}_1}{2} \log (2\pi),
\label{eq:double_m_log_prob}
\end{equation}
where $\tilde{X}_1$ is the combined input vector consisting of $X_1$ followed by $X_2$, and $\tilde{n}_1$ is the total number of observations, combining both function and derivative measurements. The predicted mean from this multi-task GP is
\begin{equation}
\bar{F}_*^{(u)} = \mu_*^{(u)}+\tilde{K}_{*1}^{(u,0)} \tilde{\Sigma}_{11}^{-1} (\tilde{Y}_1-\tilde{\mu}_1) ,
\label{eq:predict_mean}
\end{equation}
where $u$ denotes the order of the derivative and $\mu_*^{(u)}$ is obtained from a chosen mean function, which we choose to be a zero mean function.  We define kernel structures for prediction at test input(s) $X_*$ involving both function and derivative components, as
\begin{align}
\tilde{K}_{*1} &=
\begin{bmatrix} 
K_{*1} & K_{*2}^{(0,1)}
\end{bmatrix}
\\
\tilde{K}_{*1}^{(1,0)} &=
\begin{bmatrix} 
K_{*1}^{(1,0)} & K_{*2}^{(1,1)}
\end{bmatrix}
\\
\tilde{K}_{*1}^{(2,0)} &=
\begin{bmatrix} 
K_{*1}^{(2,0)} & K_{*2}^{(2,1)}
\end{bmatrix} .
\label{eq:K_tilde_star_p_p_T}
\end{align}
The covariance prediction is 
\begin{equation}
{\rm Cov}[F_*^{(u)},F_*^{(v)}] = K^{(u,v)}_{**}-\tilde{K}_{*1}^{(u,0)} \tilde{\Sigma}_{11}^{-1} \tilde{K}_{1*}^{(0,v)} .  
\end{equation}

%%%%%%%%%%%%%%%%%%%%%%%%%%%%%%%%%%
\begin{figure*}
\centering
\centering
\includegraphics[height=170pt,width=0.49\textwidth]{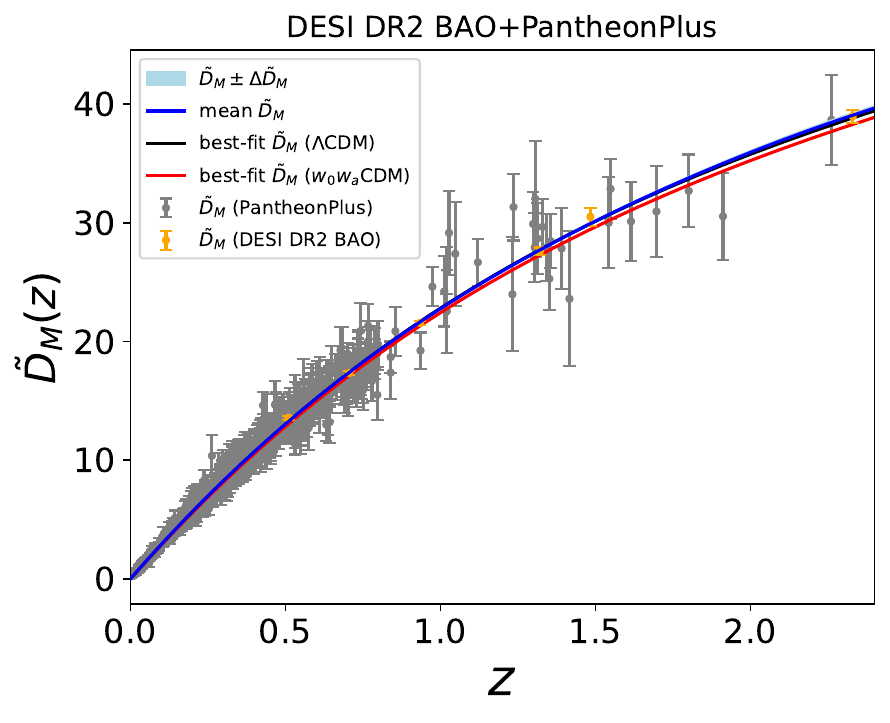}
\includegraphics[height=170pt,width=0.49\textwidth]{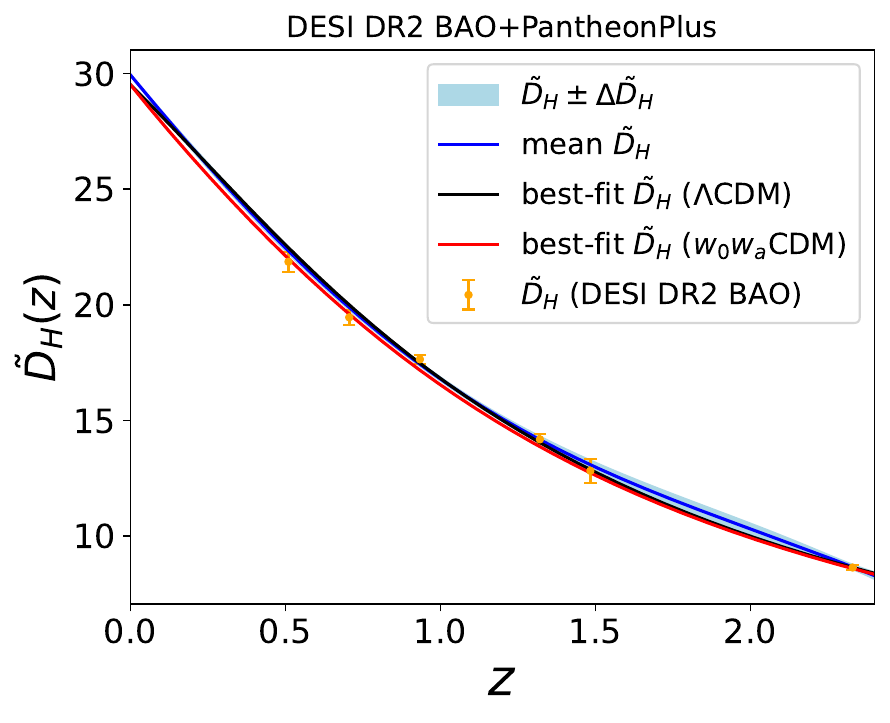} \\
\includegraphics[height=170pt,width=0.49\textwidth]{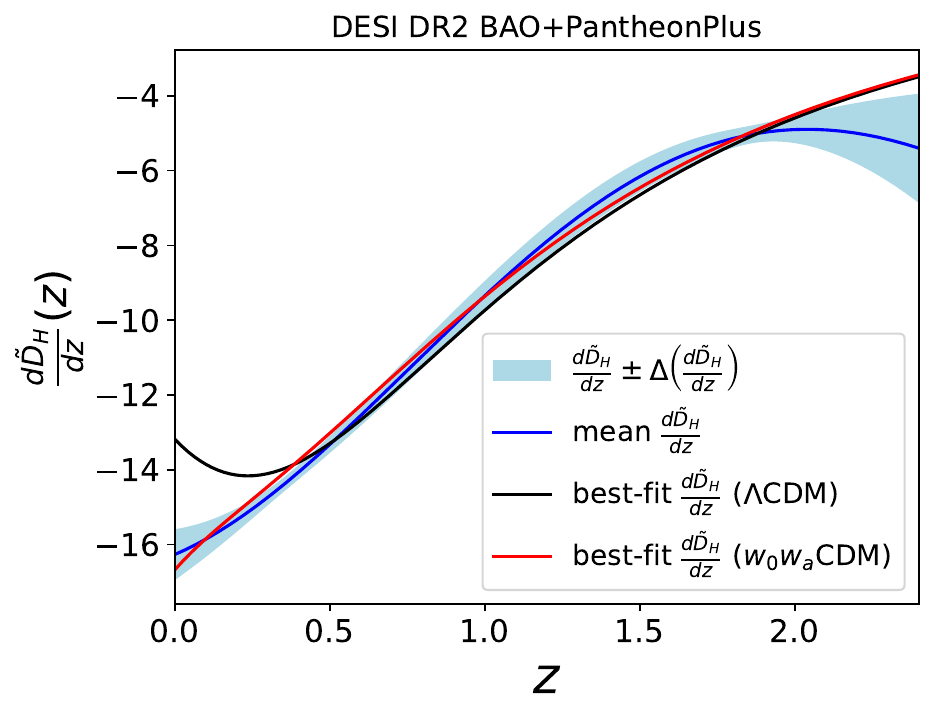}
\caption{
\label{fig:desi_pp_reconstruction}
Reconstruction of $\tilde{D}_M$ (top left), $\tilde{D}_H$ (top right), and the first (bottom left) and second (bottom right) derivatives of $\tilde{D}_H$ for the data combination DESI DR2 BAO + PantheonPlus.  Error bars are for the DESI DR2 BAO data (orange) and PantheonPlus data (grey). Solid blue lines are the reconstructed mean values obtained from the multi-task GP. Light blue shading shows the 1$\sigma$ reconstructed confidence region. Black lines are the best-fit $\Lambda$CDM model. Red lines show the best-fit $w_0w_a$CDM model.
}
\end{figure*}
%%%%%%%%%%%%%%%%%%%%%%%%%%%%%%%%%

We use multi-task GP simultaneously on BAO and Pantheon+ data to reconstruct $\tilde{D}_M$ and its derivatives. \autoref{fig:desi_pp_reconstruction} shows the reconstructed functions for the DESI DR2 BAO + PantheonPlus combination of data. The top panels show $\tilde{D}_M$ and $\tilde{D}_H$, while the bottom panels $\tilde{D}_H'$ and  $\tilde{D}_H''$. Error bars are shown for DESI DR2 BAO data (orange) and PantheonPlus data (grey). Note that PantheonPlus data actually corresponds to the observable $m_B$, but the combination of BAO + PantheonPlus data constrains the $\beta$ parameter, defined in \eqref{eq:SN_to_BAO_main}. For DESI DR2 BAO + PantheonPlus data this corresponds to $\beta=0.000519 \pm 0.000005$. With this beta, we obtain $\tilde{D}_M$ from $m_B$ for PantheonPlus data, which actually corresponds to the grey error bars in the top left panel. The solid blue lines correspond to the reconstructed mean values obtained from the predictions of multi-task GP. The light blue error regions are the reconstructed 1$\sigma$ confidence regions. Black lines show the best-fit $\Lambda$CDM model, which has $\Omega_{\rm m0}=0.312$ for the DESI DR2 BAO + PantheonPlus combination. The red lines display the best-fit $w_0w_a$CDM model, with best-fit values $\Omega_{\rm m0}=0.299$, $w_0=-0.888$, and $w_a=-0.17$ for DESI DR2 BAO + PantheonPlus data.

%%%%%%%%%%%%%%%%%%%%%%%%%%%%%%%%%%
\begin{figure*}
\centering
\centering
\includegraphics[height=170pt,width=0.49\textwidth]{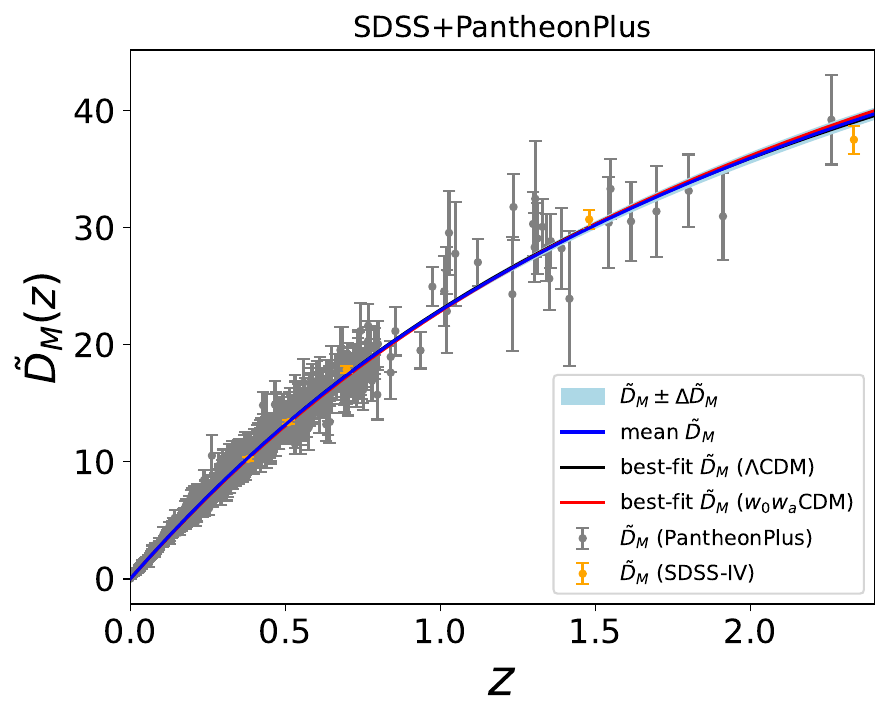}
\includegraphics[height=170pt,width=0.49\textwidth]{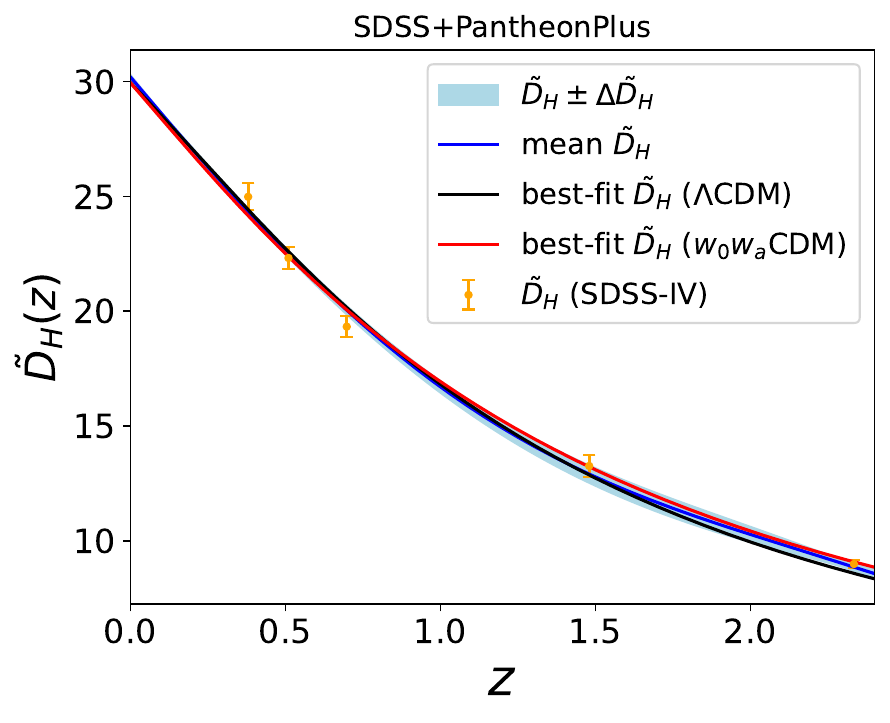} \\
\includegraphics[height=170pt,width=0.49\textwidth]{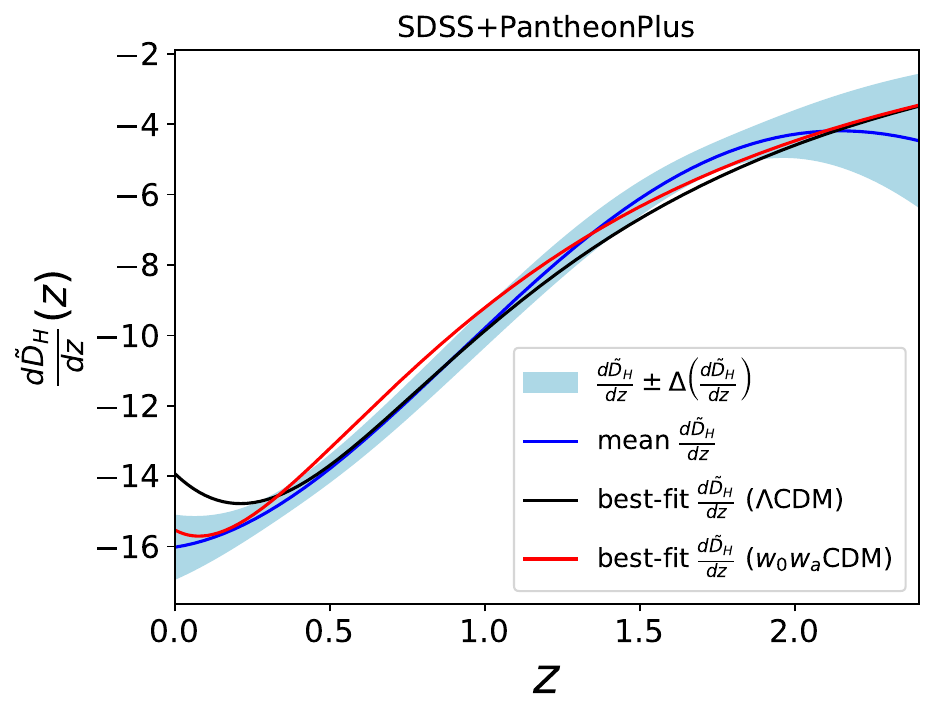}
\caption{
\label{fig:sdss_pp_reconstruction}
Same as \autoref{fig:desi_pp_reconstruction} but for SDSS + PantheonPlus data.
}
\end{figure*}
%%%%%%%%%%%%%%%%%%%%%%%%%%%%%%%%%

\autoref{fig:sdss_pp_reconstruction} follows  \autoref{fig:desi_pp_reconstruction}, but for the SDSS + PantheonPlus combination of data. Grey error bars are for the PantheonPlus data, with $\beta = 0.000526 \pm 0.000004$. The black lines show the best-fit $\Lambda$CDM model, with $\Omega_{\rm m0}=0.31$. Red lines give the best-fit  $w_0w_a$CDM model, with parameters $\Omega_{\rm m0}=0.240$, $w_0=-0.861$, and $w_a=0.30$, for SDSS + PantheonPlus data.

%%%%%%%%%%%%%%%%%%%%%%%%%%%%%%%%%%
\begin{figure*}
\centering
\centering
\includegraphics[height=170pt,width=0.49\textwidth]{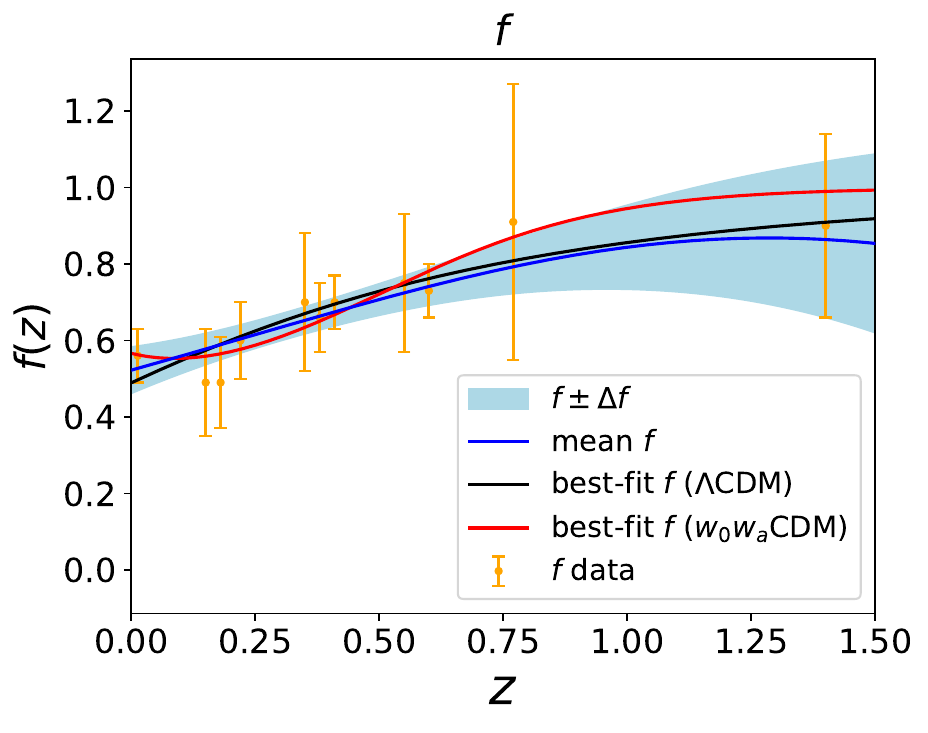}
\caption{
\label{fig:f_reconstruction}
Reconstruction of $f$. Color codes are the same as in \autoref{fig:desi_pp_reconstruction}.
}
\end{figure*}
%%%%%%%%%%%%%%%%%%%%%%%%%%%%%%%%%

\autoref{fig:f_reconstruction} presents the reconstructed smooth function of the growth rate $f$. The black line shows the best-fit $\Lambda$CDM model with $\Omega_{\rm m0}=0.276$. The red line is best-fit $w_0w_a$CDM model, with $\Omega_{\rm m0}=0.367$, $w_0=0.7$, and $w_a=-7.2$. We do not plot the reconstructed derivatives of $f$, since we do not need them for our diagnostic $G_2$. As a result, we only need a single (standard) GP instead of a multi-task GP. However, it is important to note that in FLRW space-time, any perturbation quantity is correlated with any background quantity. Thus, one can consider multi-task GP for $f$ data together with background data, such as  BAO or SNIa data. We leave this for future study. Since observational data have no correlations among them, it is a good assumption to reconstruct $f$ separately.

%%%%%%%%%%%%%%%%%%%%%%%%%%%%%%%%%%
\begin{figure*}
\centering
\centering
\includegraphics[height=170pt,width=0.49\textwidth]{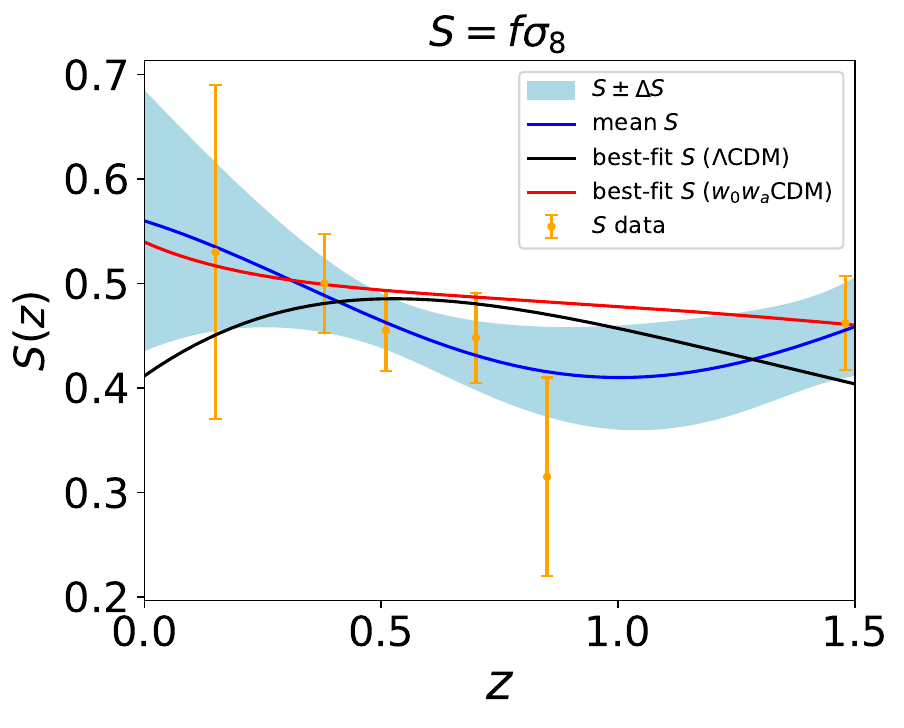}
\includegraphics[height=170pt,width=0.49\textwidth]{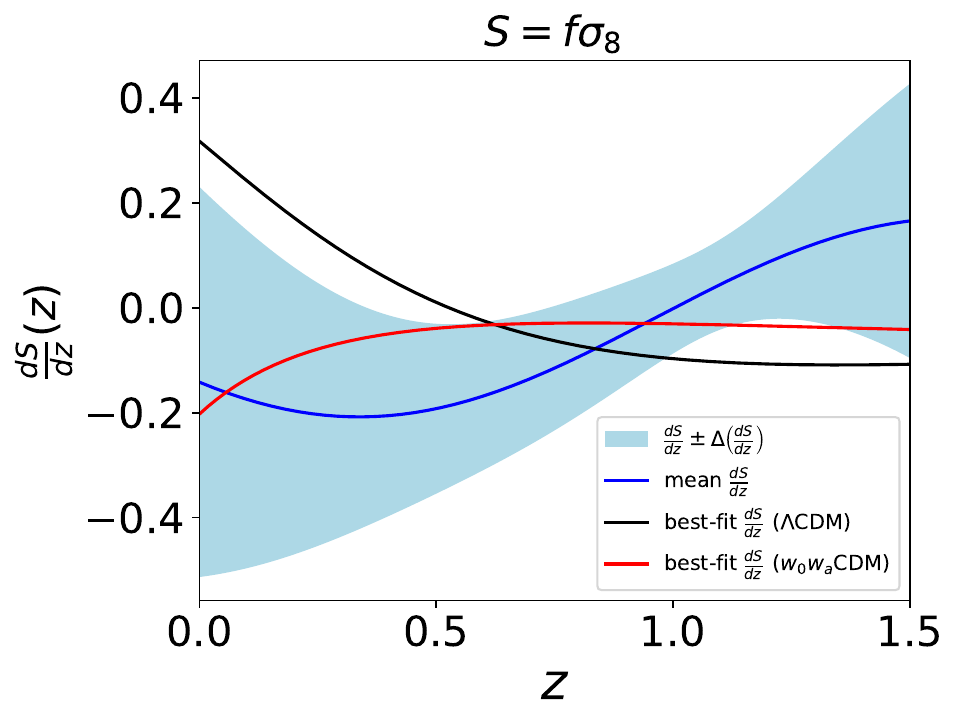}
\caption{
\label{fig:fsigma8_reconstruction}
Reconstruction of $S=f\sigma_8$ (left) and its first derivative (right). The colour codes are  as in Fig.~\ref{fig:desi_pp_reconstruction}.
}
\end{figure*}
%%%%%%%%%%%%%%%%%%%%%%%%%%%%%%%%%

In Fig.~\ref{fig:fsigma8_reconstruction}, we reconstruct the smooth function of $S=f\sigma_8$ and its first derivative. The best-fit values of the $\Lambda$CDM model parameters for the black lines are  $\Omega_{\rm m0}=0.265$ and $\sigma_{\rm 8,0}=0.861$. The red lines are for $\Omega_{\rm m0}=0.30$, $w_0=-0.04$, $w_a=-0.96$, and $\sigma_{8,0}=1.10$ for best-fit $w_0w_a$CDM model. We used standard (single-task) GP as in Fig.~\ref{fig:f_reconstruction}, for the same reason.

\section{Kernel dependence}
\label{sec-kernel}

\begin{figure*}
\centering
\centering
\includegraphics[height=170pt,width=0.49\textwidth]{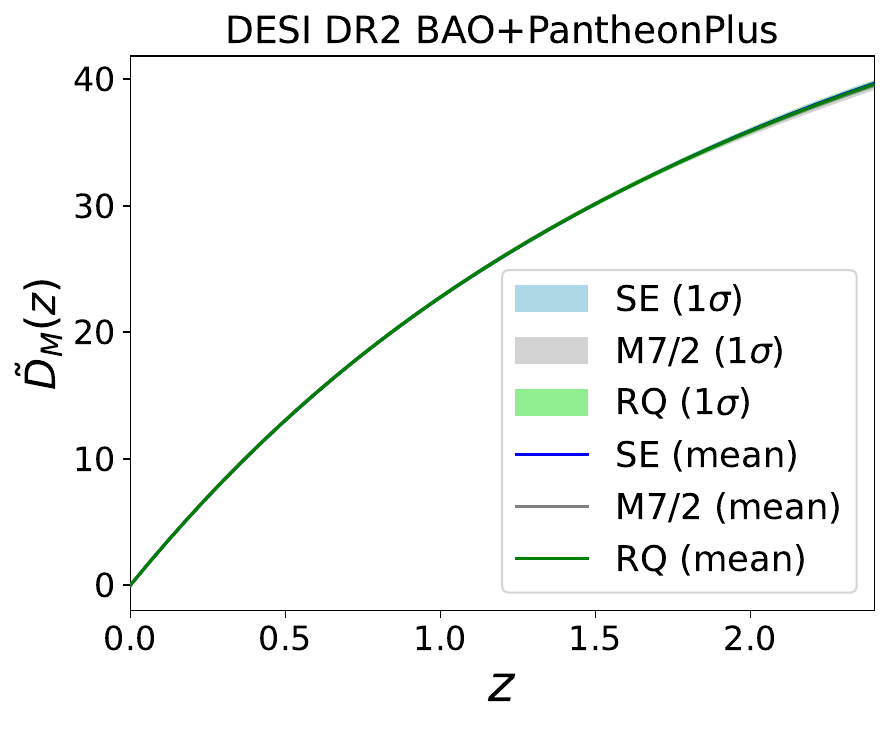}
\includegraphics[height=170pt,width=0.49\textwidth]{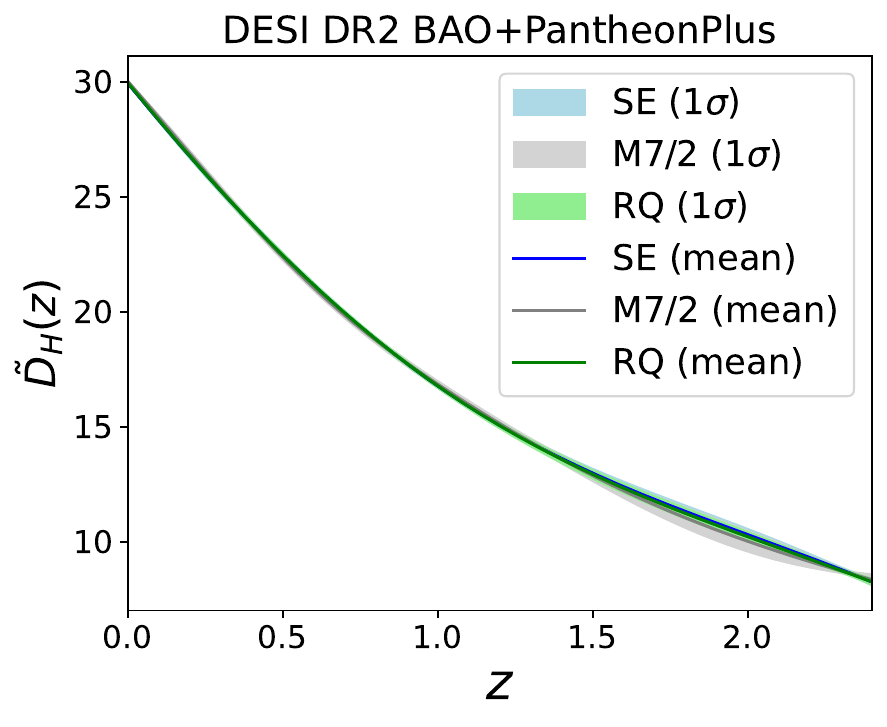} \\
\includegraphics[height=170pt,width=0.49\textwidth]{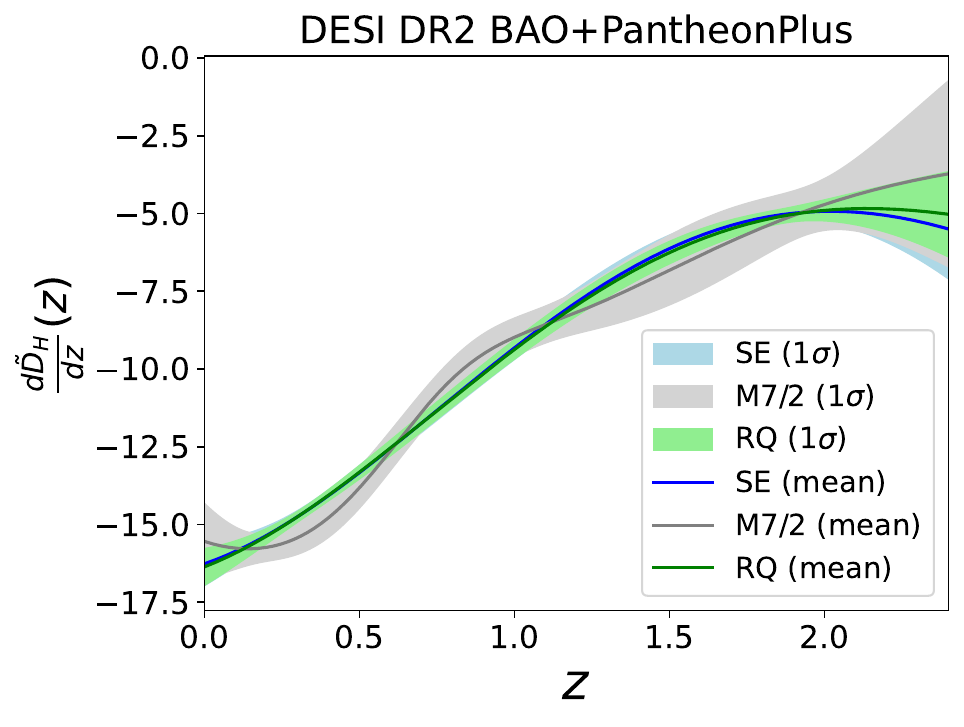}
\caption{
\label{fig:kernels}
Comparison of reconstructed functions between different kernels.
}
\end{figure*}

Here we show how the reconstructed functions depend on different kernel covariance functions in the Gaussian Processes. We compare three kernels, squared-exponential (SE), Mat\^ern 7/2 (M7/2), and rational quadratic (RQ):
\begin{eqnarray}
\text{kernel (SE)} &=& \sigma_f^2\, {\rm exp}\big[{-{d^2}/{2l^2}}\big] , 
\label{eq:SE} \\
\text{kernel (M7/2)} &=& \sigma_f^2 \left( 1+ \frac{\sqrt{7}d}{l} + \frac{14d^2}{5l^2} +\frac{7\sqrt{7}d^3}{15l^3} \right) {\rm e}^{-{\sqrt{7}d}/{l}} ,
\label{eq:MT7by2} \\
\text{kernel (RQ)} &=& \sigma_f^2 \left( 1+\frac{d^2}{2\alpha l^2} \right)^{-\alpha} ,
\label{eq:RQ}
\end{eqnarray}
where $d=|x_i-x_j|$, and other parameters are standard kernel hyper-parameters. We compare the results in \autoref{fig:kernels} and find these are quite consistent.

%%%%%%%%%%%%%%%%%%%%%%%%%%%%%%%%%%
%%%%%%%%%%%%%%%%%%%%%%%%%%%%%%%%%

\clearpage
\bibliographystyle{JHEP}
\bibliography{references}

\end{document}